\documentclass[
    aps,
    prl,
    reprint,       
    superscriptaddress, 
    amsmath,
    amssymb,
    letterpaper,   
    showpacs,      
    showkeys,      
    floatfix       
]{revtex4-2}

\usepackage{graphicx}
\usepackage{subfigure}
\usepackage{booktabs}
\usepackage{amssymb,amsmath}
\usepackage{color}
\usepackage[english]{babel}
\usepackage[utf8]{inputenc}
\usepackage[colorinlistoftodos, color=green!40, prependcaption]{todonotes}
\usepackage{amsthm}
\usepackage{mathtools}
\usepackage{physics}
\usepackage{xcolor}
\usepackage{graphicx}
\usepackage{adjustbox}
\usepackage{placeins}
\usepackage{csquotes}
\usepackage{bm}

\usepackage[pdftex, pdftitle={Article}, pdfauthor={Author}]{hyperref} 
\begin{document}
\title{Pauli `unlimited': magnetic field induced-superconductivity in UTe$_2$}

\author{Josephine J. Yu}
\affiliation{ Department of Applied Physics, Stanford University, Stanford, CA 94305, USA }
\author{Yue Yu } 
\affiliation{Department of  Physics,University of Wisconsin-Milwaukee, Milwaukee, WI 53211, USA }
\author{Chaitanya Murthy}
\affiliation{Department of Physics and Astronomy, University of Rochester, Rochester, New York 14627, USA}
\author{S. Raghu}
\affiliation{Department of Physics, Stanford University, Stanford, California 94305, USA}

\date{\today} 

\begin{abstract}
Inspired by the observation of extreme field-boosted superconductivity in uranium ditelluride, we present a scenario in which superconductivity can be {\it induced} by a strong Zeeman field, rather than destroyed by it, as is the case in Pauli-limited superconductivity.  The resulting superconducting state has an upper critical field far greater than the superconducting transition temperature, and with spin-orbit coupling, it is sensitive to the field orientation.   We demonstrate the interplay between superconductivity and metamagnetism in a simple effective theory. 
\end{abstract}

\maketitle

\textit{Introduction} --- The study of superconductivity in extreme conditions --- {\it e.g.} extreme temperature~\cite{Schilling1993}, magnetic field~\cite{Ran2019a}, or pressure~\cite{Drozdov2019} scales --- is of both fundamental and practical interest.  Progress on these issues requires conceptual advances, and a  thorough understanding of such phenomena can inform prospects for room-temperature superconductivity in real materials.  We focus here on the case of superconductivity in extreme magnetic field scales and ask how pairing remains resilient when subjected to the strong pair-breaking effects of the field.  Our analysis is inspired by the observation of superconductivity at strikingly large magnetic fields in uranium ditelluride (UTe$_2$), an unambiguous manifestation of superconductivity in extreme conditions.

In this letter, we construct an explicit scenario in which superconductivity develops at large field scales. There are two requisite conditions for the nucleation of superconductivity: a pairing `glue' and substantial density of states at the Fermi energy.  It is customary to suppose that both of these conditions must compete against deleterious effects of the field.  For instance, a Zeeman field usually acts as a pair-breaker --- this is the familiar Pauli limiting.  By contrast, we present a framework in which the  Zeeman field helps catalyze pair-formation; we are tempted to call this  Pauli `unlimited' superconductivity. While our direct inspiration is the mysterious high-field superconducting phase (HFS) of UTe$_2$, our considerations are of more general relevance to magnetic field-induced pairing in other systems.

\textit{Phenomenological considerations} --- 
For the sake of concreteness, we anchor our discussion in UTe$_2$. This system hosts multiple superconducting states at ambient pressure \cite{Ran2019,Braithwaite2019, Knafo2021a, Rosuel2022,Lewin2023,Sakai2023}, but we focus on the HFS, which is induced by magnetic fields in excess of 35 T \cite{Lewin2024, Wu2025,Helm2024, Schonemann2024}, roughly the scale associated with Kondo hybridization in this system~\cite{Ikeda2006,Aoki2013}. The complex electronic structure of UTe$_2$ \cite{Fujimori2019,Miao2020,Aoki2022,Broyles2023,Eaton2024,Weinberger2024} makes it difficult to  predict its emergent phenomena \textit{ab initio}. Instead, we analyze a vastly simplified effective theory abstracted from the most salient experimental observations. Given that the low-temperature properties of UTe$_2$ are predominantly consistent with Fermi liquid theory, we adopt this as our starting point.

A theory of the HFS in UTe$_2$ must account for several key experimental observations.  First, HFS occurs in the vicinity of metamagnetic transitions \cite{Miyake2019a, Miyake2021}, which are first order transitions as a function of magnetic field where the magnetic moment jumps abruptly.   A consistent theory must tie together both metamagnetism and HFS.  Second, HFS enjoys the feature that the scale of the upper critical field $H_{c2}$ far exceeds that of the critical temperature $T_c$. Both of these observations can be accounted for by a sizeable peak in the density of states (DOS) near the Fermi energy $E_F$.  A Zeeman field would sweep the DOS peak across the Fermi level, resulting in metamagnetism and, if sufficient attractive interactions are present, superconductivity.  If the DOS peak stems from quasiparticles with low Fermi velocity $v_F$, the resulting superconducting state will possess a short coherence length $\xi \sim v_F$
and a large upper critical field $H_{c2} \sim 1/v_F^2$.  Indeed, there is compelling evidence, both from thermodynamic \cite{Niu2020,Miyake2021, Knebel2024} and transport measurements \cite{Knafo2021a,Knebel2024}, for a large DOS being swept across $E_F$ as a function of magnetic field.

Furthermore, a theory must explain why the HFS occurs only over a range of solid angles subtended by the magnetic field \cite{Wu2025, Lewin2024}.  This observation indicates the important role played by spin-orbit coupling (SOC) in determining the magnetic response of UTe$_2$.  Thus, an effective theory of HFS must contain heavy bands with large DOS near the Fermi level; with appropriate electron correlation effects and SOC, metamagnetism and superconductivity can occur in such models and will generically depend on field orientation.  Lastly, we address the role of the magnetic field. Even in the presence of a 60 T magnetic field, UTe$_2$ remains far from the quantum limit. We thus neglect orbital effects and include only a Zeeman coupling to the field. As many of these ingredients commonly occur in heavy fermion materials, our theory may have implications for a broader set of systems.  

\textit{Model} --- In addition to capturing heavy bands and spin-orbit coupling, our model respects orthorhombic, inversion, and time-reversal (at zero-field) symmetries consistent with the crystal structure of UTe$_2$. These features are incorporated in the following Hamiltonian: 
\begin{align}
H &= H_0 + H_{\text{int}}\\
H_{0} &=  H_c +H_f + H_{c-f} +H_{\text{SOC}},
\label{eq:H0}
\end{align}
where
\begin{align}
H_c &= \sum_{\mathbf{k},\sigma,\sigma'}\left[ (\epsilon^{(c)}_\mathbf{k} - \mu_c)\delta_{\sigma\sigma'} - \mathbf{h}\cdot \pmb{\sigma}_{\sigma\sigma'} \right] c^\dagger_{\mathbf{k}\sigma}c_{\mathbf{k}\sigma'} \\
H_f &= \sum_{\mathbf{k},\sigma,\sigma'}\left[ (\epsilon^{(f)}_\mathbf{k} - \mu_f)\delta_{\sigma\sigma'} - \mathbf{h}\cdot \pmb{\sigma}_{\sigma\sigma'} \right] f^\dagger_{\mathbf{k}\sigma}f_{\mathbf{k}\sigma'}. 
\end{align}
Here, $c^{(\dagger)}_{\mathbf{k}\sigma}$ is the creation (annihilation) operator for a $c$ fermion with momentum $\mathbf{k}$ and spin $\sigma$ (similar for $f^{(\dagger)}_{\mathbf{k}\sigma}$); $\mathbf{h}$ is the applied magnetic field, $\pmb{\sigma}$ is a vector of Pauli matrices, and $\delta$ is the Kronecker delta. In UTe$_2$, the $c$ and $f$ quasiparticles are ultimately derived from Uranium 6$d$ and 5$f$ states \cite{Eaton2024}, the latter of which are neither in the local moment regime nor in the highly itinerant regime. For this reason, we do not impose a constraint on double occupancy on such states as is customary in the study of the periodic Anderson model; instead, we treat them as renormalized quasiparticle bands with low Fermi
velocity. 
We consider a three-dimensional parabolic dispersion for both species: 
\begin{equation}
    \epsilon_{\alpha}(\vec{k}) = \sum_{j=x,yz}\frac{k_j^2}{2m_{\alpha j}}.
\end{equation}
We will work in the limit that $m_{z} \gg m_{ x},m_{y}$ for both species, such that the Fermi surface and relevant properties are quasi-two-dimensional. We let the $f$ ($c$) fermions represent heavy (light) quasiparticles, so $m_{fj} \gg m_{cj}$. 

Hybridization between species is captured through
\begin{equation}
    H_{c-f} = \sum_{\mathbf{r}, \mathbf{r'},\sigma} V_{\mathbf{r},\mathbf{r'}}(c^\dagger_{\mathbf{r}\sigma}f_{\mathbf{r'}\sigma} + \text{h.c.}) 
\end{equation}
and results in bands that are admixtures of light
and heavy quasiparticles. For simplicity, we assume that the $f$ and $c$ fermions belong to orbitals with the same parity and that these orbitals transform trivially under crystal symmetry operations ({\it e.g.}~$s$-wave).  We therefore consider only on-site hybridization: $V_{\mathbf{r},\mathbf{r}'} = V_0\delta(\mathbf{r}-\mathbf{r}')$. Our qualitative conclusions are independent of these details. 

We now introduce SOC terms consistent with all symmetries and the use of $s$-wave orbitals for both $c$ and $f$:
\begin{align}
H_{SOC} &=  \sum_{n=x,y,z} \sum_{\mathbf{k},\sigma,\sigma'} \gamma_n w_n (\mathbf{k}) (-ic_{\mathbf{k}\sigma}^\dagger f_{\mathbf{k}\sigma'} \sigma_n^{\sigma\sigma'} +\text{h.c.})
\end{align}
where $w_x(\mathbf{k}) = k_y k_z$, $ w_y(\mathbf{k}) = k_x k_z$, and $w_z(\mathbf{k}) = k_x k_y$. The specific numerical values of all parameters are detailed in Appendix~\ref{app:num-param}. We describe the interactions $H_{\text{int}}$ in later sections. 

\begin{figure}[h]
\begin{centering}
\includegraphics{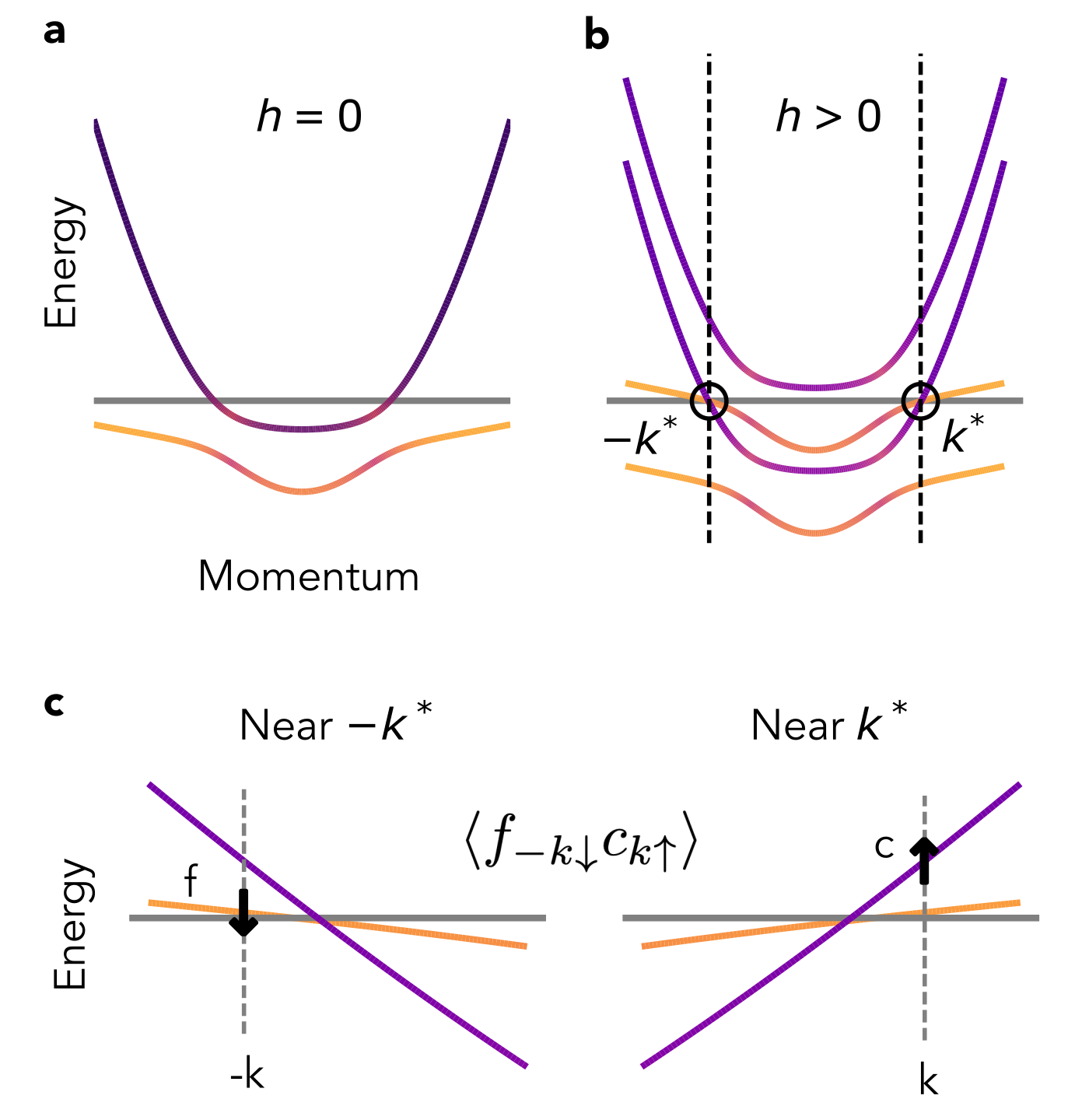}
\end{centering}
\caption{(a) Schematic of hybridized bands at zero-field ($h=0$). (b) Spin-split bands in the presence of an applied field ($h>0$), dashed lines mark locations of crossings (denoted $k^*$ and $-k^*$) between opposite-spin bands at the Fermi level (gray line). (c) Close-up view of regions near $k^*$ and $-k^*$, where one band has mostly $f$-character (orange) and the other has mostly $c$-character (purple). Arrows mark an example of Cooper pairing between the heavy spin-down (orange) and light spin-up (purple) quasiparticles at $\pm k$.
\label{fig:cartoon}}
\end{figure}

\textit{Field-induced superconductivity} --- We now illustrate how field-induced superconductivity arises in this model. The field scale at which superconductivity appears in UTe$_2$ is comparable to the Kondo scale, suggesting that both spin species are close to the Fermi energy and potentially relevant for superconductivity. We thus propose a plausible scenario for pairing in high field which involves both spin species through inter-band pairing. 

The primary agent is the Zeeman effect, which 1) splits the zero-field bands into (pseudo)spin-polarized bands and 2) allows for band crossings (Fig.~\ref{fig:cartoon}a, b). In the limit that the Zeeman field and hybridization dominate over SOC ($h>V_0\gg \gamma_n$), the opposite-spin quasiparticle bands generically have well-defined species character at the crossing (Fig.~\ref{fig:cartoon}c). Then, any attraction involving opposite species and opposite spin, such as antiferromagnetic, transverse spin, or phonon-mediated density-density interactions can induce pairing. Since one band generically has low Fermi velocity, it has a broad set of available states at this crossing in $k$-space, thus enhancing the tendency towards uniform $(k,-k)$ pairing regardless of the form of the dispersion of the other band. This distinguishes our proposal from previous studies of field-induced uniform superconductivity \cite{Salamone2023} and pair-density waves \cite{Han2022, Chakraborty2024, Clepkens2024} arising from a similar mechanism of field-induced band crossing.

As illustrated in Fig.~\ref{fig:cartoon}c, in the simplest case, the pairing is momentum-independent: $\Delta_0 \equiv \expval{f_{-k \downarrow} c_{k \uparrow}}$. This $\Delta_0$ pairing is a generalization of the conventional spin-singlet; it is an even-parity state, antisymmetric under exchange of combined species and spin. Like a conventional spin-singlet state, the $\Delta_0$ state has no symmetry-protected nodes. Similarly, a conventional spin singlet is destroyed in a magnetic field strong enough to traverse the spin gap (``Pauli limited"), and $\Delta_0$ is unsurprisingly subject to this same limit, as there is a maximum field strength beyond which the bands crossing near the Fermi level separate. However, $\Delta_0$ differs from the conventional spin singlet state in that it is also ``Pauli \textit{un}limited" --- a minimum field strength is required to induce the bands to cross near the Fermi energy. Thus, the Zeeman field acts as both a pair-breaker and a pair-maker in our model, which suggests why superconductivity occurs only in a range of magnetic field strengths. 

\textit{Pair susceptibility} --- The above arguments are captured quantitatively in the pair susceptibility:
\label{sec:susc}
\begin{align}
\chi_{ij}(q)= \frac{1}{\beta} \frac{1}{N}\sum_{\omega_n, k} \text{Tr} \Big[G(k,i\omega_n) D_i^\dagger G(-k+q,-i\omega_n) D_j  \Big],
\label{eq:susc}
\end{align}
where $G$ is the $4\times4$ matrix of fermion propagators and $D_i$ are $4\times4$ pairing vertices. The quantity $\chi_{ij}$ represents the susceptibility of the normal state to pair in the $i$-th superconducting channel if subject to a pair field with the form $D_j$.   The largest eigenvalue $\lambda_{\text{max}}$ of the matrix $\chi_{ij}$ is a proxy for the superconducting transition temperature $T_c$ and the ``strength'' of pairing. 

\begin{figure}[h]
    \centering
    \includegraphics{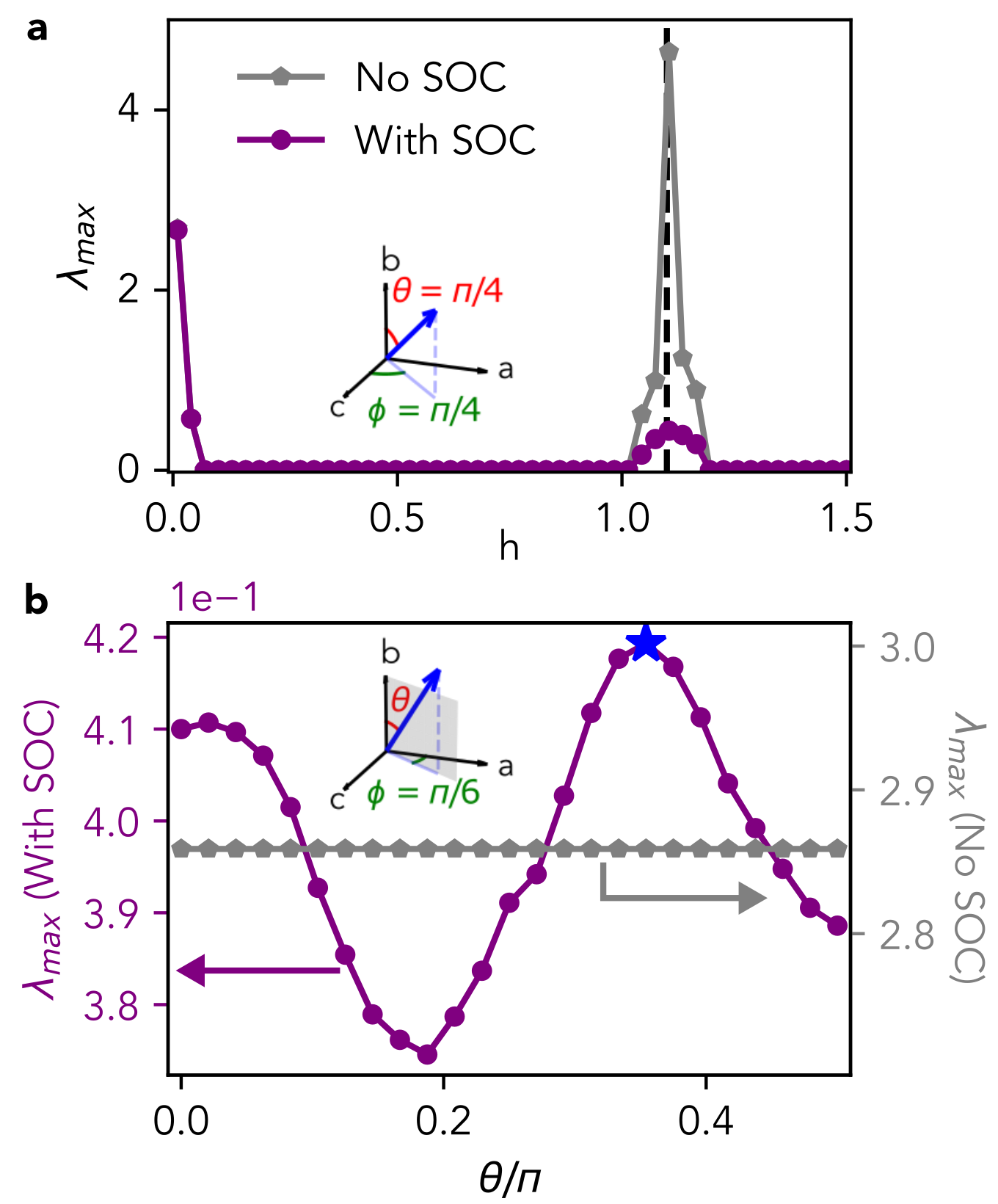}
    \caption{Largest eigenvalue $\lambda_{\text{max}}$ of the pair susceptibility matrix $\chi$ as a function of: (a) field magnitude $h$ for the field at a fixed orientation of $\phi=\theta=\pi/4$ and (b) angle $\theta$ for fixed magnitude ($h=1.1$) and $\phi= \pi/6$. Purple circles indicate data with SOC, grey pentagons indicate data without SOC. The blue star indicates the largest value of $\lambda_{\text{max}}$ for the field in the plane $\phi=\pi/6$ shown in (b); this is the direction of the field ($\phi=\pi/6$, $\theta \approx7\pi/20$) for which we obtain mean-field results (Fig.~\ref{fig:mf}). The insets show the orientation of the magnetic field. Note the difference in left and right $y$-axis scales for plot (b). }
    \label{fig:susc} 
\end{figure}

To explore the inter-band pairing proposed previously, we consider even-parity gap structures $D_i$ which involve opposite species (Appendix~\ref{app:ham-matrix}). We focus on zero center-of-mass momentum ($q=0$) pairing, and a discussion of finite-$q$ pairing can be found in Appendix~\ref{app:finite-q}.  

The largest eigenvalue $\lambda_{\text{max}}$ of the pair susceptibility matrix $\chi_{ij}(q=0)$ is shown in Fig.~\ref{fig:susc} as a function of both field magnitude (at a fixed orientation) and field orientation (at a fixed magnitude).  As a function of field magnitude (oriented at polar angle $\theta = \pi/4$ with respect to the $b$ axis and azimuthal angle $\phi=\pi/4$), the largest eigenvalue is finite at zero-field but is quickly suppressed due to usual Pauli-limiting (Fig.~\ref{fig:susc}a).  At larger field strengths, pairing is again enhanced as two bands intersect at the Fermi level. Qualitatively, the presence of SOC generates small mismatches between the intersecting bands (avoided crossings), which reduces but does not eliminate the field-induced pairing. 

The susceptibility is necessarily isotropic in the absence of SOC, as shown in Fig.~\ref{fig:susc}b (grey pentagons) for a representative range of orientations (other orientations are shown in Appendix~\ref{app:chi-orientation}). If SOC is included (purple circles), then $\lambda_{\text{max}}$ becomes highly anisotropic; at a fixed azimuthal angle $\phi$,  $\lambda_{\text{max}}$ is maximized at an arbitrary polar angle $\theta$ (not necessarily along a crystalline axis). This anisotropy arises from the competition and cooperation among the three different SOC terms at a given field strength; for certain field directions, the splitting due to SOC is minimal (maximal), resulting in local maxima (minima) of $\chi$ as a function of orientation.  

While the pair susceptibility analysis suggests the field-dependence of superconductivity, it is most reliable when the susceptibility diverges. As illustrated in Fig.~\ref{fig:cartoon}, the Fermi velocities of the two bands are different at the crossing, and this suppresses the familiar ``Cooper logarithm'' usually responsible for a diverging susceptibility in conventional, weak-coupling BCS theory. To obtain superconductivity, we thus require finite attraction, so our theory is closer to the Stoner theory of magnetism~\cite{Stoner1997} than the BCS theory of superconductivity~\cite{Bardeen1957}. 

\textit{Mean-field analysis} --- To complete our analysis, we must incorporate interactions. As a consequence, we can now observe the interplay between metamagnetism and superconductivity.  Though metamagnetism does not drive the superconductivity in our model, we demonstrate that it arises naturally at the mean-field level. We include repulsive on-site Hubbard interactions 
\begin{equation}
    H_U = \sum_{\alpha=c,f} U_\alpha\sum_{i} n_{i\alpha\uparrow}n_{i\alpha\downarrow},
\end{equation}
with $U_\alpha>0$, which decouple in the magnetization channel. In practice, we will take $U_f\gg U_c$ such that the interactions are stronger in the flatter band.  

We must also introduce an attractive interaction, which determines the form of resulting superconductivity. We consider a few possible sources of attraction. The simplest such source is from a phonon-mediated density-density interaction, which is generically present in all crystalline solids. There is also evidence for rich magnetic behavior in UTe$_2$~\cite{Knafo2021, Butch2022,Tokunaga2023, Wu2025}; in principle, there could be ferromagnetic, antiferromagnetic, and transverse spin interactions. For simplicity, we focus on a single interaction type, an attractive density-density interaction:
\begin{equation}
    H_g =-g\sum_{i}(n_{ic} +n_{if})^2 ,
\end{equation}
with $g>0$, which favors superconductivity of the opposite-species, opposite-spin type described previously. This choice is one of convenience, and considering instead antiferromagnetic or transverse spin interactions produces similar results (Appendix~\ref{app:gap-struct}).

We treat the interactions $H_U$ and $H_g$ in a mean-field approximation by defining the mean fields for magnetization $\mathbf{M}_\alpha$ for each species $\alpha=c,f$ and the superconducting gaps $\Delta_i$ in each pairing channel (Appendix~\ref{app:MFT}).  By minimizing the total free energy (Appendix~\ref{app:minimization}) as a function of $\mathbf{M}_\alpha$ and $\Delta_i$, we simultaneously find solutions for the magnetizations and gap magnitudes. 

\begin{figure}
\begin{centering}
\includegraphics{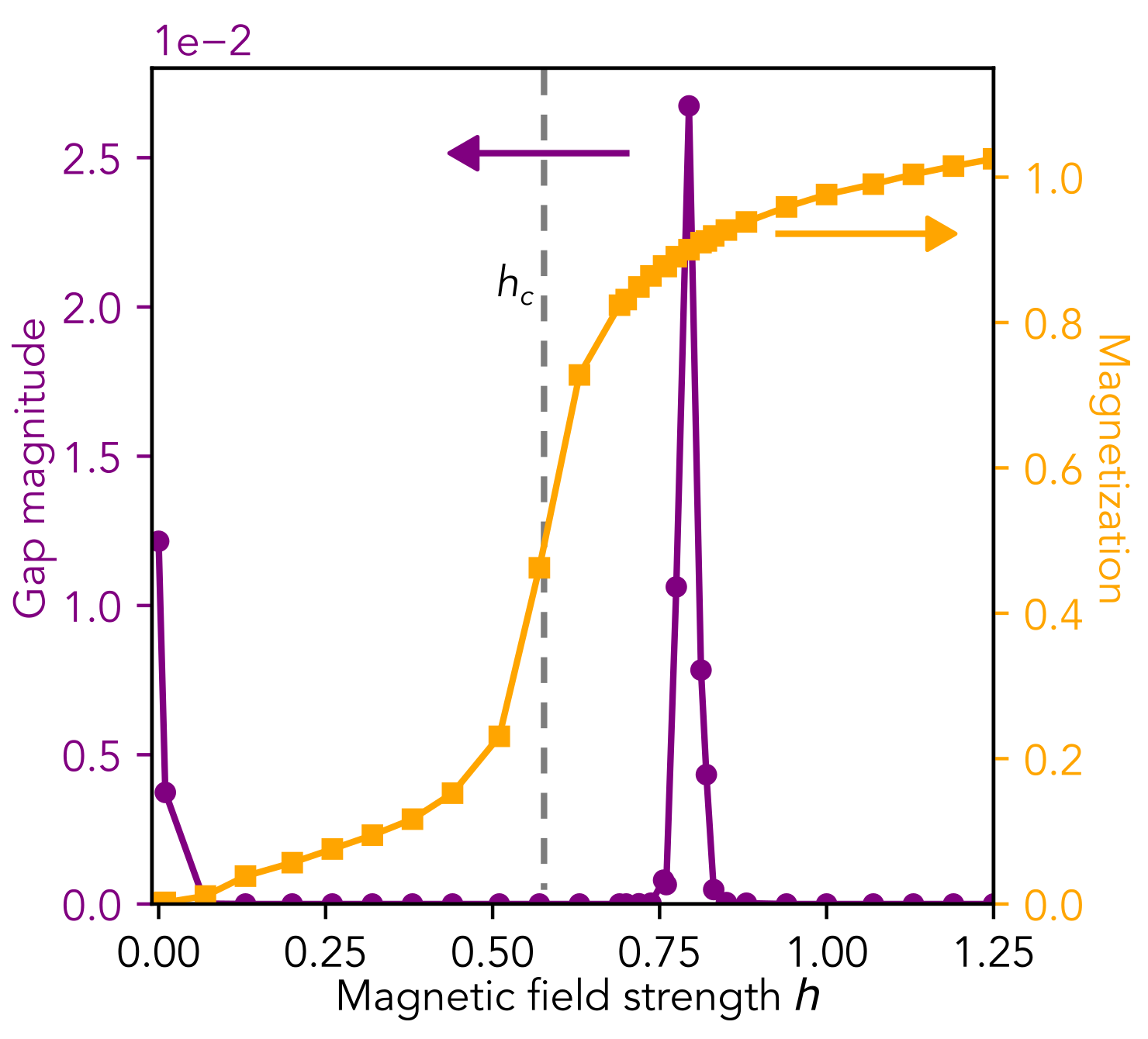}
\end{centering}
\caption{Results of minimizing the free energy of the mean-field Hamiltonian (Appendix~\ref{app:minimization}) with $g=4$, $U_f=1$ for a magnetic field $h$ in the orientation $\phi=\pi/3$, $\theta = 0.35\pi$ (orange star in Fig. \ref{fig:susc}c). Left axis: maximum gap magnitude at each magnetic field strength. Right axis: total magnetization at each magnetic field strength. Critical field for the metamagnetic transition $h_c$ is marked by a dashed line.}
\label{fig:mf}
\end{figure}

The results of this mean-field analysis are shown in Fig.~\ref{fig:mf} as a function of field strength $h$ (for fixed field orientation). There is a range of field strengths (approximately $h=0.75$ to $h=0.8$) in which the superconducting gap has a finite magnitude. This field-induced gap is unsurprisingly non-unitary (Appendix~\ref{app:gap-nonunitary}). We also find a metamagnetic transition, at which the magnitude of the total magnetization $\mathbf{M} = \mathbf{M}_c +\mathbf{M}_f$ increases abruptly at a critical field strength $h_c$; this transition sharpens with increasing Hubbard interaction $U_f$. In Fig.~\ref{fig:mf}, superconductivity is induced at field strengths greater than $h_c$; more generally, the band parameters control whether superconductivity onsets at a field strength at or above $h_c$ (Appendix~\ref{app:MFT-relative-field}). The mean-field results are a concrete demonstration of how inter-band superconductivity can be induced by a field and how superconductivity and metamagnetism can appear together.  

\textit{Odd-parity pairing} --- Thus far, we have focused on even-parity, inter-band pairing. However, a natural candidate for superconductivity in a large magnetic field is odd-parity, non-unitary spin-triplet pairing (such as that proposed in \cite{Lewin2024}), since it is not subject to usual Pauli limiting. Within our model, such pairing can be induced by nearest-neighbor ferromagnetic interactions, for example. If the pairing primarily occurs on the heavy $f$ band, then superconductivity will again arise in a range of finite magnetic fields strengths; a minimum field strength is required to bring the spin-down $f$ band to the Fermi level, and at some maximum field strength, this band is completely depopulated. In this scenario, the range of field strengths at which superconductivity nucleates is necessarily the same as that over which the metamagnetic transition occurs (Appendix~\ref{app:odd-parity}). 

\textit{Implications for experiment} --- While experiments have observed an enhancement of the Sommerfeld coefficient $\gamma$ of heat capacity and a peak in the $A$-coefficient of resistivity at the metamagnetic transition for a field $\mathbf{h} \parallel \hat{b}$, such measurements for other field orientations would further bolster our perspective that a band with a large DOS near the Fermi energy is essential. We have also invoked SOC to generate the field orientation-dependence of the superconductivity. In this picture, the gap magnitudes inherit the anisotropy, and $T_c$ would depend on the field orientation, as has been experimentally observed for a portion of the HFS \cite{Wu2025a}. Lastly, in the case of even-parity pairing, the relationship between the field strength at which superconductivity occurs and the strength of the metamagnetic transition can be controlled by changing band parameters. In experiment, pressure can tune these effective parameters; indeed, a separation of the field scale for the metamagnetic transition and the onset of superconductivity has been observed under pressure \cite{Ran2021}.    

In summary, we presented a scenario for field-induced superconductivity, inspired by the observation of HFS in UTe$_2$.  We focused on a Zeeman-induced (``Pauli unlimited"), interband pairing state and commented on the possibility of intraband odd-parity pairing. Our work offers a conceptual foothold for the study of both orbital field effects and the role of quenched randomness in the high-field superconducting phase.  We relegate the study of these questions to future work. Our work is founded on a simplified Fermi liquid model, but we aim to forge a connection between this effective theory and more microscopic descriptions of UTe$_2$ in future work.  While our analysis here has largely been inspired by the phenonemology of UTe$_2$, the essential ingredients for field-induced pairing are sufficiently general; we conjecture that similar field-induced superconductivity may lurk in a broader array of quantum materials.

\section*{Acknowledgments} 
We thank E. Berg, S. Brown, N. Butch, A. Eaton, M. Grosche, S. Lewin, J. Paglione, B. Ramshaw, P. Rosa, C. Varma,  and the participants of the Aspen Center for Physics Winter 2025 conference on Unconventional and High-Temperature Superconductivity for helpful discussions.  J.J.Y. and S.R. are supported in part by the US Department of Energy, Office of Basic Energy Sciences, Division of Materials Sciences and Engineering, under Contract No. DE-AC02-76SF00515. 

\bibliography{bibliography.bib}

\begin{thebibliography}{37}%
\makeatletter
\providecommand \@ifxundefined [1]{%
 \@ifx{#1\undefined}
}%
\providecommand \@ifnum [1]{%
 \ifnum #1\expandafter \@firstoftwo
 \else \expandafter \@secondoftwo
 \fi
}%
\providecommand \@ifx [1]{%
 \ifx #1\expandafter \@firstoftwo
 \else \expandafter \@secondoftwo
 \fi
}%
\providecommand \natexlab [1]{#1}%
\providecommand \enquote  [1]{``#1''}%
\providecommand \bibnamefont  [1]{#1}%
\providecommand \bibfnamefont [1]{#1}%
\providecommand \citenamefont [1]{#1}%
\providecommand \href@noop [0]{\@secondoftwo}%
\providecommand \href [0]{\begingroup \@sanitize@url \@href}%
\providecommand \@href[1]{\@@startlink{#1}\@@href}%
\providecommand \@@href[1]{\endgroup#1\@@endlink}%
\providecommand \@sanitize@url [0]{\catcode `\\12\catcode `\$12\catcode `\&12\catcode `\#12\catcode `\^12\catcode `\_12\catcode `\%12\relax}%
\providecommand \@@startlink[1]{}%
\providecommand \@@endlink[0]{}%
\providecommand \url  [0]{\begingroup\@sanitize@url \@url }%
\providecommand \@url [1]{\endgroup\@href {#1}{\urlprefix }}%
\providecommand \urlprefix  [0]{URL }%
\providecommand \Eprint [0]{\href }%
\providecommand \doibase [0]{https://doi.org/}%
\providecommand \selectlanguage [0]{\@gobble}%
\providecommand \bibinfo  [0]{\@secondoftwo}%
\providecommand \bibfield  [0]{\@secondoftwo}%
\providecommand \translation [1]{[#1]}%
\providecommand \BibitemOpen [0]{}%
\providecommand \bibitemStop [0]{}%
\providecommand \bibitemNoStop [0]{.\EOS\space}%
\providecommand \EOS [0]{\spacefactor3000\relax}%
\providecommand \BibitemShut  [1]{\csname bibitem#1\endcsname}%
\let\auto@bib@innerbib\@empty
\bibitem [{\citenamefont {Schilling}\ \emph {et~al.}(1993)\citenamefont {Schilling}, \citenamefont {Cantoni}, \citenamefont {Guo},\ and\ \citenamefont {Ott}}]{Schilling1993}%
  \BibitemOpen
  \bibfield  {author} {\bibinfo {author} {\bibfnamefont {A.}~\bibnamefont {Schilling}}, \bibinfo {author} {\bibfnamefont {M.}~\bibnamefont {Cantoni}}, \bibinfo {author} {\bibfnamefont {J.}~\bibnamefont {Guo}},\ and\ \bibinfo {author} {\bibfnamefont {H.}~\bibnamefont {Ott}},\ }\href@noop {} {\bibfield  {journal} {\bibinfo  {journal} {Nature}\ }\textbf {\bibinfo {volume} {363}},\ \bibinfo {pages} {56} (\bibinfo {year} {1993})}\BibitemShut {NoStop}%
\bibitem [{\citenamefont {Ran}\ \emph {et~al.}(2019{\natexlab{a}})\citenamefont {Ran}, \citenamefont {Liu}, \citenamefont {Eo}, \citenamefont {Campbell}, \citenamefont {Neves}, \citenamefont {Fuhrman}, \citenamefont {Saha}, \citenamefont {Eckberg}, \citenamefont {Kim}, \citenamefont {Graf} \emph {et~al.}}]{Ran2019a}%
  \BibitemOpen
  \bibfield  {author} {\bibinfo {author} {\bibfnamefont {S.}~\bibnamefont {Ran}}, \bibinfo {author} {\bibfnamefont {I.-L.}\ \bibnamefont {Liu}}, \bibinfo {author} {\bibfnamefont {Y.~S.}\ \bibnamefont {Eo}}, \bibinfo {author} {\bibfnamefont {D.~J.}\ \bibnamefont {Campbell}}, \bibinfo {author} {\bibfnamefont {P.~M.}\ \bibnamefont {Neves}}, \bibinfo {author} {\bibfnamefont {W.~T.}\ \bibnamefont {Fuhrman}}, \bibinfo {author} {\bibfnamefont {S.~R.}\ \bibnamefont {Saha}}, \bibinfo {author} {\bibfnamefont {C.}~\bibnamefont {Eckberg}}, \bibinfo {author} {\bibfnamefont {H.}~\bibnamefont {Kim}}, \bibinfo {author} {\bibfnamefont {D.}~\bibnamefont {Graf}}, \emph {et~al.},\ }\href@noop {} {\bibfield  {journal} {\bibinfo  {journal} {Nature physics}\ }\textbf {\bibinfo {volume} {15}},\ \bibinfo {pages} {1250} (\bibinfo {year} {2019}{\natexlab{a}})}\BibitemShut {NoStop}%
\bibitem [{\citenamefont {Drozdov}\ \emph {et~al.}(2019)\citenamefont {Drozdov}, \citenamefont {Kong}, \citenamefont {Minkov}, \citenamefont {Besedin}, \citenamefont {Kuzovnikov}, \citenamefont {Mozaffari}, \citenamefont {Balicas}, \citenamefont {Balakirev}, \citenamefont {Graf}, \citenamefont {Prakapenka} \emph {et~al.}}]{Drozdov2019}%
  \BibitemOpen
  \bibfield  {author} {\bibinfo {author} {\bibfnamefont {A.}~\bibnamefont {Drozdov}}, \bibinfo {author} {\bibfnamefont {P.}~\bibnamefont {Kong}}, \bibinfo {author} {\bibfnamefont {V.}~\bibnamefont {Minkov}}, \bibinfo {author} {\bibfnamefont {S.}~\bibnamefont {Besedin}}, \bibinfo {author} {\bibfnamefont {M.}~\bibnamefont {Kuzovnikov}}, \bibinfo {author} {\bibfnamefont {S.}~\bibnamefont {Mozaffari}}, \bibinfo {author} {\bibfnamefont {L.}~\bibnamefont {Balicas}}, \bibinfo {author} {\bibfnamefont {F.~F.}\ \bibnamefont {Balakirev}}, \bibinfo {author} {\bibfnamefont {D.}~\bibnamefont {Graf}}, \bibinfo {author} {\bibfnamefont {V.}~\bibnamefont {Prakapenka}}, \emph {et~al.},\ }\href@noop {} {\bibfield  {journal} {\bibinfo  {journal} {Nature}\ }\textbf {\bibinfo {volume} {569}},\ \bibinfo {pages} {528} (\bibinfo {year} {2019})}\BibitemShut {NoStop}%
\bibitem [{\citenamefont {Ran}\ \emph {et~al.}(2019{\natexlab{b}})\citenamefont {Ran}, \citenamefont {Eckberg}, \citenamefont {Ding}, \citenamefont {Furukawa}, \citenamefont {Metz}, \citenamefont {Saha}, \citenamefont {Liu}, \citenamefont {Zic}, \citenamefont {Kim}, \citenamefont {Paglione},\ and\ \citenamefont {Butch}}]{Ran2019}%
  \BibitemOpen
  \bibfield  {author} {\bibinfo {author} {\bibfnamefont {S.}~\bibnamefont {Ran}}, \bibinfo {author} {\bibfnamefont {C.}~\bibnamefont {Eckberg}}, \bibinfo {author} {\bibfnamefont {Q.-P.}\ \bibnamefont {Ding}}, \bibinfo {author} {\bibfnamefont {Y.}~\bibnamefont {Furukawa}}, \bibinfo {author} {\bibfnamefont {T.}~\bibnamefont {Metz}}, \bibinfo {author} {\bibfnamefont {S.~R.}\ \bibnamefont {Saha}}, \bibinfo {author} {\bibfnamefont {I.-L.}\ \bibnamefont {Liu}}, \bibinfo {author} {\bibfnamefont {M.}~\bibnamefont {Zic}}, \bibinfo {author} {\bibfnamefont {H.}~\bibnamefont {Kim}}, \bibinfo {author} {\bibfnamefont {J.}~\bibnamefont {Paglione}},\ and\ \bibinfo {author} {\bibfnamefont {N.~P.}\ \bibnamefont {Butch}},\ }\href {https://doi.org/10.1126/science.aav8645} {\bibfield  {journal} {\bibinfo  {journal} {Science}\ }\textbf {\bibinfo {volume} {365}},\ \bibinfo {pages} {684} (\bibinfo {year} {2019}{\natexlab{b}})}\BibitemShut {NoStop}%
\bibitem [{\citenamefont {Braithwaite}\ \emph {et~al.}(2019)\citenamefont {Braithwaite}, \citenamefont {Vali{\v s}ka}, \citenamefont {Knebel}, \citenamefont {Lapertot}, \citenamefont {Brison}, \citenamefont {Pourret}, \citenamefont {Zhitomirsky}, \citenamefont {Flouquet}, \citenamefont {Honda},\ and\ \citenamefont {Aoki}}]{Braithwaite2019}%
  \BibitemOpen
  \bibfield  {author} {\bibinfo {author} {\bibfnamefont {D.}~\bibnamefont {Braithwaite}}, \bibinfo {author} {\bibfnamefont {M.}~\bibnamefont {Vali{\v s}ka}}, \bibinfo {author} {\bibfnamefont {G.}~\bibnamefont {Knebel}}, \bibinfo {author} {\bibfnamefont {G.}~\bibnamefont {Lapertot}}, \bibinfo {author} {\bibfnamefont {J.-P.}\ \bibnamefont {Brison}}, \bibinfo {author} {\bibfnamefont {A.}~\bibnamefont {Pourret}}, \bibinfo {author} {\bibfnamefont {M.~E.}\ \bibnamefont {Zhitomirsky}}, \bibinfo {author} {\bibfnamefont {J.}~\bibnamefont {Flouquet}}, \bibinfo {author} {\bibfnamefont {F.}~\bibnamefont {Honda}},\ and\ \bibinfo {author} {\bibfnamefont {D.}~\bibnamefont {Aoki}},\ }\href {https://doi.org/10.1038/s42005-019-0248-z} {\bibfield  {journal} {\bibinfo  {journal} {Communications Physics}\ }\textbf {\bibinfo {volume} {2}},\ \bibinfo {pages} {1} (\bibinfo {year} {2019})}\BibitemShut {NoStop}%
\bibitem [{\citenamefont {Knafo}\ \emph {et~al.}(2021{\natexlab{a}})\citenamefont {Knafo}, \citenamefont {Nardone}, \citenamefont {Vali{\v s}ka}, \citenamefont {Zitouni}, \citenamefont {Lapertot}, \citenamefont {Aoki}, \citenamefont {Knebel},\ and\ \citenamefont {Braithwaite}}]{Knafo2021a}%
  \BibitemOpen
  \bibfield  {author} {\bibinfo {author} {\bibfnamefont {W.}~\bibnamefont {Knafo}}, \bibinfo {author} {\bibfnamefont {M.}~\bibnamefont {Nardone}}, \bibinfo {author} {\bibfnamefont {M.}~\bibnamefont {Vali{\v s}ka}}, \bibinfo {author} {\bibfnamefont {A.}~\bibnamefont {Zitouni}}, \bibinfo {author} {\bibfnamefont {G.}~\bibnamefont {Lapertot}}, \bibinfo {author} {\bibfnamefont {D.}~\bibnamefont {Aoki}}, \bibinfo {author} {\bibfnamefont {G.}~\bibnamefont {Knebel}},\ and\ \bibinfo {author} {\bibfnamefont {D.}~\bibnamefont {Braithwaite}},\ }\href {https://doi.org/10.1038/s42005-021-00545-z} {\bibfield  {journal} {\bibinfo  {journal} {Communications Physics}\ }\textbf {\bibinfo {volume} {4}},\ \bibinfo {pages} {1} (\bibinfo {year} {2021}{\natexlab{a}})}\BibitemShut {NoStop}%
\bibitem [{\citenamefont {Rosuel}\ \emph {et~al.}(2022)\citenamefont {Rosuel}, \citenamefont {Marcenat}, \citenamefont {Knebel}, \citenamefont {Klein}, \citenamefont {Pourret}, \citenamefont {Marquardt}, \citenamefont {Niu}, \citenamefont {Rousseau}, \citenamefont {Demuer}, \citenamefont {Seyfarth}, \citenamefont {Lapertot}, \citenamefont {Aoki}, \citenamefont {Braithwaite}, \citenamefont {Flouquet},\ and\ \citenamefont {Brison}}]{Rosuel2022}%
  \BibitemOpen
  \bibfield  {author} {\bibinfo {author} {\bibfnamefont {A.}~\bibnamefont {Rosuel}}, \bibinfo {author} {\bibfnamefont {C.}~\bibnamefont {Marcenat}}, \bibinfo {author} {\bibfnamefont {G.}~\bibnamefont {Knebel}}, \bibinfo {author} {\bibfnamefont {T.}~\bibnamefont {Klein}}, \bibinfo {author} {\bibfnamefont {A.}~\bibnamefont {Pourret}}, \bibinfo {author} {\bibfnamefont {N.}~\bibnamefont {Marquardt}}, \bibinfo {author} {\bibfnamefont {Q.}~\bibnamefont {Niu}}, \bibinfo {author} {\bibfnamefont {S.}~\bibnamefont {Rousseau}}, \bibinfo {author} {\bibfnamefont {A.}~\bibnamefont {Demuer}}, \bibinfo {author} {\bibfnamefont {G.}~\bibnamefont {Seyfarth}}, \bibinfo {author} {\bibfnamefont {G.}~\bibnamefont {Lapertot}}, \bibinfo {author} {\bibfnamefont {D.}~\bibnamefont {Aoki}}, \bibinfo {author} {\bibfnamefont {D.}~\bibnamefont {Braithwaite}}, \bibinfo {author} {\bibfnamefont {J.}~\bibnamefont {Flouquet}},\ and\ \bibinfo {author} {\bibfnamefont {J.-P.}\ \bibnamefont {Brison}},\ }\href@noop {} {\bibinfo {title} {Field-induced
  tuning of the pairing state in a superconductor}} (\bibinfo {year} {2022}),\ \Eprint {https://arxiv.org/abs/2205.04524} {arXiv:2205.04524 [cond-mat]} \BibitemShut {NoStop}%
\bibitem [{\citenamefont {Lewin}\ \emph {et~al.}(2023)\citenamefont {Lewin}, \citenamefont {Frank}, \citenamefont {Ran}, \citenamefont {Paglione},\ and\ \citenamefont {Butch}}]{Lewin2023}%
  \BibitemOpen
  \bibfield  {author} {\bibinfo {author} {\bibfnamefont {S.~K.}\ \bibnamefont {Lewin}}, \bibinfo {author} {\bibfnamefont {C.~E.}\ \bibnamefont {Frank}}, \bibinfo {author} {\bibfnamefont {S.}~\bibnamefont {Ran}}, \bibinfo {author} {\bibfnamefont {J.}~\bibnamefont {Paglione}},\ and\ \bibinfo {author} {\bibfnamefont {N.~P.}\ \bibnamefont {Butch}},\ }\href {https://doi.org/10.1088/1361-6633/acfb93} {\bibfield  {journal} {\bibinfo  {journal} {Reports on Progress in Physics}\ }\textbf {\bibinfo {volume} {86}},\ \bibinfo {pages} {114501} (\bibinfo {year} {2023})}\BibitemShut {NoStop}%
\bibitem [{\citenamefont {Sakai}\ \emph {et~al.}(2023)\citenamefont {Sakai}, \citenamefont {Tokiwa}, \citenamefont {Opletal}, \citenamefont {Kimata}, \citenamefont {Awaji}, \citenamefont {Sasaki}, \citenamefont {Aoki}, \citenamefont {Kambe}, \citenamefont {Tokunaga},\ and\ \citenamefont {Haga}}]{Sakai2023}%
  \BibitemOpen
  \bibfield  {author} {\bibinfo {author} {\bibfnamefont {H.}~\bibnamefont {Sakai}}, \bibinfo {author} {\bibfnamefont {Y.}~\bibnamefont {Tokiwa}}, \bibinfo {author} {\bibfnamefont {P.}~\bibnamefont {Opletal}}, \bibinfo {author} {\bibfnamefont {M.}~\bibnamefont {Kimata}}, \bibinfo {author} {\bibfnamefont {S.}~\bibnamefont {Awaji}}, \bibinfo {author} {\bibfnamefont {T.}~\bibnamefont {Sasaki}}, \bibinfo {author} {\bibfnamefont {D.}~\bibnamefont {Aoki}}, \bibinfo {author} {\bibfnamefont {S.}~\bibnamefont {Kambe}}, \bibinfo {author} {\bibfnamefont {Y.}~\bibnamefont {Tokunaga}},\ and\ \bibinfo {author} {\bibfnamefont {Y.}~\bibnamefont {Haga}},\ }\href {https://doi.org/10.1103/PhysRevLett.130.196002} {\bibfield  {journal} {\bibinfo  {journal} {Physical Review Letters}\ }\textbf {\bibinfo {volume} {130}},\ \bibinfo {pages} {196002} (\bibinfo {year} {2023})}\BibitemShut {NoStop}%
\bibitem [{\citenamefont {Lewin}\ \emph {et~al.}(2024)\citenamefont {Lewin}, \citenamefont {Czajka}, \citenamefont {Frank}, \citenamefont {Salas}, \citenamefont {Yoon}, \citenamefont {Eo}, \citenamefont {Paglione}, \citenamefont {Nevidomskyy}, \citenamefont {Singleton},\ and\ \citenamefont {Butch}}]{Lewin2024}%
  \BibitemOpen
  \bibfield  {author} {\bibinfo {author} {\bibfnamefont {S.~K.}\ \bibnamefont {Lewin}}, \bibinfo {author} {\bibfnamefont {P.}~\bibnamefont {Czajka}}, \bibinfo {author} {\bibfnamefont {C.~E.}\ \bibnamefont {Frank}}, \bibinfo {author} {\bibfnamefont {G.~S.}\ \bibnamefont {Salas}}, \bibinfo {author} {\bibfnamefont {H.}~\bibnamefont {Yoon}}, \bibinfo {author} {\bibfnamefont {Y.~S.}\ \bibnamefont {Eo}}, \bibinfo {author} {\bibfnamefont {J.}~\bibnamefont {Paglione}}, \bibinfo {author} {\bibfnamefont {A.~H.}\ \bibnamefont {Nevidomskyy}}, \bibinfo {author} {\bibfnamefont {J.}~\bibnamefont {Singleton}},\ and\ \bibinfo {author} {\bibfnamefont {N.~P.}\ \bibnamefont {Butch}},\ }\href@noop {} {\bibinfo {title} {High-{{Field Superconducting Halo}} in {{UTe}}\$\_2\$}} (\bibinfo {year} {2024}),\ \Eprint {https://arxiv.org/abs/2402.18564} {arXiv:2402.18564 [cond-mat]} \BibitemShut {NoStop}%
\bibitem [{\citenamefont {Wu}\ \emph {et~al.}(2025{\natexlab{a}})\citenamefont {Wu}, \citenamefont {Weinberger}, \citenamefont {Hickey}, \citenamefont {Chichinadze}, \citenamefont {Shaffer}, \citenamefont {Cabala}, \citenamefont {Chen}, \citenamefont {Long}, \citenamefont {Brumm}, \citenamefont {Xie}, \citenamefont {Lin}, \citenamefont {Skourski}, \citenamefont {Zhu}, \citenamefont {Graf}, \citenamefont {Sechovsky}, \citenamefont {Lonzarich}, \citenamefont {Valiska}, \citenamefont {Grosche},\ and\ \citenamefont {Eaton}}]{Wu2025}%
  \BibitemOpen
  \bibfield  {author} {\bibinfo {author} {\bibfnamefont {Z.}~\bibnamefont {Wu}}, \bibinfo {author} {\bibfnamefont {T.~I.}\ \bibnamefont {Weinberger}}, \bibinfo {author} {\bibfnamefont {A.~J.}\ \bibnamefont {Hickey}}, \bibinfo {author} {\bibfnamefont {D.~V.}\ \bibnamefont {Chichinadze}}, \bibinfo {author} {\bibfnamefont {D.}~\bibnamefont {Shaffer}}, \bibinfo {author} {\bibfnamefont {A.}~\bibnamefont {Cabala}}, \bibinfo {author} {\bibfnamefont {H.}~\bibnamefont {Chen}}, \bibinfo {author} {\bibfnamefont {M.}~\bibnamefont {Long}}, \bibinfo {author} {\bibfnamefont {T.~J.}\ \bibnamefont {Brumm}}, \bibinfo {author} {\bibfnamefont {W.}~\bibnamefont {Xie}}, \bibinfo {author} {\bibfnamefont {Y.}~\bibnamefont {Lin}}, \bibinfo {author} {\bibfnamefont {Y.}~\bibnamefont {Skourski}}, \bibinfo {author} {\bibfnamefont {Z.}~\bibnamefont {Zhu}}, \bibinfo {author} {\bibfnamefont {D.~E.}\ \bibnamefont {Graf}}, \bibinfo {author} {\bibfnamefont {V.}~\bibnamefont {Sechovsky}}, \bibinfo {author} {\bibfnamefont {G.~G.}\ \bibnamefont
  {Lonzarich}}, \bibinfo {author} {\bibfnamefont {M.}~\bibnamefont {Valiska}}, \bibinfo {author} {\bibfnamefont {F.~M.}\ \bibnamefont {Grosche}},\ and\ \bibinfo {author} {\bibfnamefont {A.~G.}\ \bibnamefont {Eaton}},\ }\href {https://doi.org/10.48550/arXiv.2403.02535} {\bibinfo {title} {A quantum critical line bounds the high field metamagnetic transition surface in {{UTe}}\$\_2\$}} (\bibinfo {year} {2025}{\natexlab{a}}),\ \Eprint {https://arxiv.org/abs/2403.02535} {arXiv:2403.02535 [cond-mat]} \BibitemShut {NoStop}%
\bibitem [{\citenamefont {Helm}\ \emph {et~al.}(2024)\citenamefont {Helm}, \citenamefont {Kimata}, \citenamefont {Sudo}, \citenamefont {Miyata}, \citenamefont {Stirnat}, \citenamefont {F{\"o}rster}, \citenamefont {Hornung}, \citenamefont {K{\"o}nig}, \citenamefont {Sheikin}, \citenamefont {Pourret}, \citenamefont {Lapertot}, \citenamefont {Aoki}, \citenamefont {Knebel}, \citenamefont {Wosnitza},\ and\ \citenamefont {Brison}}]{Helm2024}%
  \BibitemOpen
  \bibfield  {author} {\bibinfo {author} {\bibfnamefont {T.}~\bibnamefont {Helm}}, \bibinfo {author} {\bibfnamefont {M.}~\bibnamefont {Kimata}}, \bibinfo {author} {\bibfnamefont {K.}~\bibnamefont {Sudo}}, \bibinfo {author} {\bibfnamefont {A.}~\bibnamefont {Miyata}}, \bibinfo {author} {\bibfnamefont {J.}~\bibnamefont {Stirnat}}, \bibinfo {author} {\bibfnamefont {T.}~\bibnamefont {F{\"o}rster}}, \bibinfo {author} {\bibfnamefont {J.}~\bibnamefont {Hornung}}, \bibinfo {author} {\bibfnamefont {M.}~\bibnamefont {K{\"o}nig}}, \bibinfo {author} {\bibfnamefont {I.}~\bibnamefont {Sheikin}}, \bibinfo {author} {\bibfnamefont {A.}~\bibnamefont {Pourret}}, \bibinfo {author} {\bibfnamefont {G.}~\bibnamefont {Lapertot}}, \bibinfo {author} {\bibfnamefont {D.}~\bibnamefont {Aoki}}, \bibinfo {author} {\bibfnamefont {G.}~\bibnamefont {Knebel}}, \bibinfo {author} {\bibfnamefont {J.}~\bibnamefont {Wosnitza}},\ and\ \bibinfo {author} {\bibfnamefont {J.-P.}\ \bibnamefont {Brison}},\ }\href
  {https://doi.org/10.1038/s41467-023-44183-1} {\bibfield  {journal} {\bibinfo  {journal} {Nature Communications}\ }\textbf {\bibinfo {volume} {15}},\ \bibinfo {pages} {37} (\bibinfo {year} {2024})},\ \Eprint {https://arxiv.org/abs/2207.08261} {arXiv:2207.08261 [cond-mat]} \BibitemShut {NoStop}%
\bibitem [{\citenamefont {Sch{\"o}nemann}\ \emph {et~al.}(2024)\citenamefont {Sch{\"o}nemann}, \citenamefont {Rosa}, \citenamefont {Thomas}, \citenamefont {Lai}, \citenamefont {Nguyen}, \citenamefont {Singleton}, \citenamefont {Brosha}, \citenamefont {McDonald}, \citenamefont {Zapf}, \citenamefont {Maiorov},\ and\ \citenamefont {Jaime}}]{Schonemann2024}%
  \BibitemOpen
  \bibfield  {author} {\bibinfo {author} {\bibfnamefont {R.}~\bibnamefont {Sch{\"o}nemann}}, \bibinfo {author} {\bibfnamefont {P.~F.~S.}\ \bibnamefont {Rosa}}, \bibinfo {author} {\bibfnamefont {S.~M.}\ \bibnamefont {Thomas}}, \bibinfo {author} {\bibfnamefont {Y.}~\bibnamefont {Lai}}, \bibinfo {author} {\bibfnamefont {D.~N.}\ \bibnamefont {Nguyen}}, \bibinfo {author} {\bibfnamefont {J.}~\bibnamefont {Singleton}}, \bibinfo {author} {\bibfnamefont {E.~L.}\ \bibnamefont {Brosha}}, \bibinfo {author} {\bibfnamefont {R.~D.}\ \bibnamefont {McDonald}}, \bibinfo {author} {\bibfnamefont {V.}~\bibnamefont {Zapf}}, \bibinfo {author} {\bibfnamefont {B.}~\bibnamefont {Maiorov}},\ and\ \bibinfo {author} {\bibfnamefont {M.}~\bibnamefont {Jaime}},\ }\href {https://doi.org/10.1093/pnasnexus/pgad428} {\bibfield  {journal} {\bibinfo  {journal} {PNAS Nexus}\ }\textbf {\bibinfo {volume} {3}},\ \bibinfo {pages} {pgad428} (\bibinfo {year} {2024})}\BibitemShut {NoStop}%
\bibitem [{\citenamefont {Ikeda}\ \emph {et~al.}(2006)\citenamefont {Ikeda}, \citenamefont {Sakai}, \citenamefont {Aoki}, \citenamefont {Homma}, \citenamefont {Yamamoto}, \citenamefont {Nakamura}, \citenamefont {Shiokawa}, \citenamefont {Haga},\ and\ \citenamefont {{\=O}nuki}}]{Ikeda2006}%
  \BibitemOpen
  \bibfield  {author} {\bibinfo {author} {\bibfnamefont {S.}~\bibnamefont {Ikeda}}, \bibinfo {author} {\bibfnamefont {H.}~\bibnamefont {Sakai}}, \bibinfo {author} {\bibfnamefont {D.}~\bibnamefont {Aoki}}, \bibinfo {author} {\bibfnamefont {Y.}~\bibnamefont {Homma}}, \bibinfo {author} {\bibfnamefont {E.}~\bibnamefont {Yamamoto}}, \bibinfo {author} {\bibfnamefont {A.}~\bibnamefont {Nakamura}}, \bibinfo {author} {\bibfnamefont {Y.}~\bibnamefont {Shiokawa}}, \bibinfo {author} {\bibfnamefont {Y.}~\bibnamefont {Haga}},\ and\ \bibinfo {author} {\bibfnamefont {Y.}~\bibnamefont {{\=O}nuki}},\ }\href {https://doi.org/10.1143/JPSJS.75S.116} {\bibfield  {journal} {\bibinfo  {journal} {Journal of the Physical Society of Japan}\ }\textbf {\bibinfo {volume} {75}},\ \bibinfo {pages} {116} (\bibinfo {year} {2006})}\BibitemShut {NoStop}%
\bibitem [{\citenamefont {Aoki}\ \emph {et~al.}(2013)\citenamefont {Aoki}, \citenamefont {Knafo},\ and\ \citenamefont {Sheikin}}]{Aoki2013}%
  \BibitemOpen
  \bibfield  {author} {\bibinfo {author} {\bibfnamefont {D.}~\bibnamefont {Aoki}}, \bibinfo {author} {\bibfnamefont {W.}~\bibnamefont {Knafo}},\ and\ \bibinfo {author} {\bibfnamefont {I.}~\bibnamefont {Sheikin}},\ }\href@noop {} {\bibfield  {journal} {\bibinfo  {journal} {Comptes Rendus Physique}\ }\textbf {\bibinfo {volume} {14}},\ \bibinfo {pages} {53} (\bibinfo {year} {2013})}\BibitemShut {NoStop}%
\bibitem [{\citenamefont {Fujimori}\ \emph {et~al.}(2019)\citenamefont {Fujimori}, \citenamefont {Kawasaki}, \citenamefont {Takeda}, \citenamefont {Yamagami}, \citenamefont {Nakamura}, \citenamefont {Homma},\ and\ \citenamefont {Aoki}}]{Fujimori2019}%
  \BibitemOpen
  \bibfield  {author} {\bibinfo {author} {\bibfnamefont {S.-i.}\ \bibnamefont {Fujimori}}, \bibinfo {author} {\bibfnamefont {I.}~\bibnamefont {Kawasaki}}, \bibinfo {author} {\bibfnamefont {Y.}~\bibnamefont {Takeda}}, \bibinfo {author} {\bibfnamefont {H.}~\bibnamefont {Yamagami}}, \bibinfo {author} {\bibfnamefont {A.}~\bibnamefont {Nakamura}}, \bibinfo {author} {\bibfnamefont {Y.}~\bibnamefont {Homma}},\ and\ \bibinfo {author} {\bibfnamefont {D.}~\bibnamefont {Aoki}},\ }\href {https://doi.org/10.7566/JPSJ.88.103701} {\bibfield  {journal} {\bibinfo  {journal} {Journal of the Physical Society of Japan}\ }\textbf {\bibinfo {volume} {88}},\ \bibinfo {pages} {103701} (\bibinfo {year} {2019})}\BibitemShut {NoStop}%
\bibitem [{\citenamefont {Miao}\ \emph {et~al.}(2020)\citenamefont {Miao}, \citenamefont {Liu}, \citenamefont {Xu}, \citenamefont {Kotta}, \citenamefont {Kang}, \citenamefont {Ran}, \citenamefont {Paglione}, \citenamefont {Kotliar}, \citenamefont {Butch}, \citenamefont {Denlinger},\ and\ \citenamefont {Wray}}]{Miao2020}%
  \BibitemOpen
  \bibfield  {author} {\bibinfo {author} {\bibfnamefont {L.}~\bibnamefont {Miao}}, \bibinfo {author} {\bibfnamefont {S.}~\bibnamefont {Liu}}, \bibinfo {author} {\bibfnamefont {Y.}~\bibnamefont {Xu}}, \bibinfo {author} {\bibfnamefont {E.~C.}\ \bibnamefont {Kotta}}, \bibinfo {author} {\bibfnamefont {C.-J.}\ \bibnamefont {Kang}}, \bibinfo {author} {\bibfnamefont {S.}~\bibnamefont {Ran}}, \bibinfo {author} {\bibfnamefont {J.}~\bibnamefont {Paglione}}, \bibinfo {author} {\bibfnamefont {G.}~\bibnamefont {Kotliar}}, \bibinfo {author} {\bibfnamefont {N.~P.}\ \bibnamefont {Butch}}, \bibinfo {author} {\bibfnamefont {J.~D.}\ \bibnamefont {Denlinger}},\ and\ \bibinfo {author} {\bibfnamefont {L.~A.}\ \bibnamefont {Wray}},\ }\href {https://doi.org/10.1103/PhysRevLett.124.076401} {\bibfield  {journal} {\bibinfo  {journal} {Physical Review Letters}\ }\textbf {\bibinfo {volume} {124}},\ \bibinfo {pages} {076401} (\bibinfo {year} {2020})}\BibitemShut {NoStop}%
\bibitem [{\citenamefont {Aoki}\ \emph {et~al.}(2022)\citenamefont {Aoki}, \citenamefont {Sakai}, \citenamefont {Opletal}, \citenamefont {Tokiwa}, \citenamefont {Ishizuka}, \citenamefont {Yanase}, \citenamefont {Harima}, \citenamefont {Nakamura}, \citenamefont {Li}, \citenamefont {Homma}, \citenamefont {Shimizu}, \citenamefont {Knebel}, \citenamefont {Flouquet},\ and\ \citenamefont {Haga}}]{Aoki2022}%
  \BibitemOpen
  \bibfield  {author} {\bibinfo {author} {\bibfnamefont {D.}~\bibnamefont {Aoki}}, \bibinfo {author} {\bibfnamefont {H.}~\bibnamefont {Sakai}}, \bibinfo {author} {\bibfnamefont {P.}~\bibnamefont {Opletal}}, \bibinfo {author} {\bibfnamefont {Y.}~\bibnamefont {Tokiwa}}, \bibinfo {author} {\bibfnamefont {J.}~\bibnamefont {Ishizuka}}, \bibinfo {author} {\bibfnamefont {Y.}~\bibnamefont {Yanase}}, \bibinfo {author} {\bibfnamefont {H.}~\bibnamefont {Harima}}, \bibinfo {author} {\bibfnamefont {A.}~\bibnamefont {Nakamura}}, \bibinfo {author} {\bibfnamefont {D.}~\bibnamefont {Li}}, \bibinfo {author} {\bibfnamefont {Y.}~\bibnamefont {Homma}}, \bibinfo {author} {\bibfnamefont {Y.}~\bibnamefont {Shimizu}}, \bibinfo {author} {\bibfnamefont {G.}~\bibnamefont {Knebel}}, \bibinfo {author} {\bibfnamefont {J.}~\bibnamefont {Flouquet}},\ and\ \bibinfo {author} {\bibfnamefont {Y.}~\bibnamefont {Haga}},\ }\href {https://doi.org/10.7566/JPSJ.91.083704} {\bibfield  {journal} {\bibinfo  {journal} {Journal of the Physical Society of
  Japan}\ }\textbf {\bibinfo {volume} {91}},\ \bibinfo {pages} {083704} (\bibinfo {year} {2022})}\BibitemShut {NoStop}%
\bibitem [{\citenamefont {Broyles}\ \emph {et~al.}(2023)\citenamefont {Broyles}, \citenamefont {Rehfuss}, \citenamefont {Siddiquee}, \citenamefont {Zhu}, \citenamefont {Zheng}, \citenamefont {Nikolo}, \citenamefont {Graf}, \citenamefont {Singleton},\ and\ \citenamefont {Ran}}]{Broyles2023}%
  \BibitemOpen
  \bibfield  {author} {\bibinfo {author} {\bibfnamefont {C.}~\bibnamefont {Broyles}}, \bibinfo {author} {\bibfnamefont {Z.}~\bibnamefont {Rehfuss}}, \bibinfo {author} {\bibfnamefont {H.}~\bibnamefont {Siddiquee}}, \bibinfo {author} {\bibfnamefont {J.~A.}\ \bibnamefont {Zhu}}, \bibinfo {author} {\bibfnamefont {K.}~\bibnamefont {Zheng}}, \bibinfo {author} {\bibfnamefont {M.}~\bibnamefont {Nikolo}}, \bibinfo {author} {\bibfnamefont {D.}~\bibnamefont {Graf}}, \bibinfo {author} {\bibfnamefont {J.}~\bibnamefont {Singleton}},\ and\ \bibinfo {author} {\bibfnamefont {S.}~\bibnamefont {Ran}},\ }\href {https://doi.org/10.1103/PhysRevLett.131.036501} {\bibfield  {journal} {\bibinfo  {journal} {Physical Review Letters}\ }\textbf {\bibinfo {volume} {131}},\ \bibinfo {pages} {036501} (\bibinfo {year} {2023})}\BibitemShut {NoStop}%
\bibitem [{\citenamefont {Eaton}\ \emph {et~al.}(2024)\citenamefont {Eaton}, \citenamefont {Weinberger}, \citenamefont {Popiel}, \citenamefont {Wu}, \citenamefont {Hickey}, \citenamefont {Cabala}, \citenamefont {Posp{\'i}{\v s}il}, \citenamefont {Prokle{\v s}ka}, \citenamefont {Haidamak}, \citenamefont {Bastien}, \citenamefont {Opletal}, \citenamefont {Sakai}, \citenamefont {Haga}, \citenamefont {Nowell}, \citenamefont {Benjamin}, \citenamefont {Sechovsk{\'y}}, \citenamefont {Lonzarich}, \citenamefont {Grosche},\ and\ \citenamefont {Vali{\v s}ka}}]{Eaton2024}%
  \BibitemOpen
  \bibfield  {author} {\bibinfo {author} {\bibfnamefont {A.~G.}\ \bibnamefont {Eaton}}, \bibinfo {author} {\bibfnamefont {T.~I.}\ \bibnamefont {Weinberger}}, \bibinfo {author} {\bibfnamefont {N.~J.~M.}\ \bibnamefont {Popiel}}, \bibinfo {author} {\bibfnamefont {Z.}~\bibnamefont {Wu}}, \bibinfo {author} {\bibfnamefont {A.~J.}\ \bibnamefont {Hickey}}, \bibinfo {author} {\bibfnamefont {A.}~\bibnamefont {Cabala}}, \bibinfo {author} {\bibfnamefont {J.}~\bibnamefont {Posp{\'i}{\v s}il}}, \bibinfo {author} {\bibfnamefont {J.}~\bibnamefont {Prokle{\v s}ka}}, \bibinfo {author} {\bibfnamefont {T.}~\bibnamefont {Haidamak}}, \bibinfo {author} {\bibfnamefont {G.}~\bibnamefont {Bastien}}, \bibinfo {author} {\bibfnamefont {P.}~\bibnamefont {Opletal}}, \bibinfo {author} {\bibfnamefont {H.}~\bibnamefont {Sakai}}, \bibinfo {author} {\bibfnamefont {Y.}~\bibnamefont {Haga}}, \bibinfo {author} {\bibfnamefont {R.}~\bibnamefont {Nowell}}, \bibinfo {author} {\bibfnamefont {S.~M.}\ \bibnamefont {Benjamin}}, \bibinfo {author}
  {\bibfnamefont {V.}~\bibnamefont {Sechovsk{\'y}}}, \bibinfo {author} {\bibfnamefont {G.~G.}\ \bibnamefont {Lonzarich}}, \bibinfo {author} {\bibfnamefont {F.~M.}\ \bibnamefont {Grosche}},\ and\ \bibinfo {author} {\bibfnamefont {M.}~\bibnamefont {Vali{\v s}ka}},\ }\href {https://doi.org/10.1038/s41467-023-44110-4} {\bibfield  {journal} {\bibinfo  {journal} {Nature Communications}\ }\textbf {\bibinfo {volume} {15}},\ \bibinfo {pages} {223} (\bibinfo {year} {2024})}\BibitemShut {NoStop}%
\bibitem [{\citenamefont {Weinberger}\ \emph {et~al.}(2024)\citenamefont {Weinberger}, \citenamefont {Wu}, \citenamefont {Graf}, \citenamefont {Skourski}, \citenamefont {Cabala}, \citenamefont {Posp{\'i}{\v s}il}, \citenamefont {Prokle{\v s}ka}, \citenamefont {Haidamak}, \citenamefont {Bastien}, \citenamefont {Sechovsk{\'y}}, \citenamefont {Lonzarich}, \citenamefont {Vali{\v s}ka}, \citenamefont {Grosche},\ and\ \citenamefont {Eaton}}]{Weinberger2024}%
  \BibitemOpen
  \bibfield  {author} {\bibinfo {author} {\bibfnamefont {T.~I.}\ \bibnamefont {Weinberger}}, \bibinfo {author} {\bibfnamefont {Z.}~\bibnamefont {Wu}}, \bibinfo {author} {\bibfnamefont {D.~E.}\ \bibnamefont {Graf}}, \bibinfo {author} {\bibfnamefont {Y.}~\bibnamefont {Skourski}}, \bibinfo {author} {\bibfnamefont {A.}~\bibnamefont {Cabala}}, \bibinfo {author} {\bibfnamefont {J.}~\bibnamefont {Posp{\'i}{\v s}il}}, \bibinfo {author} {\bibfnamefont {J.}~\bibnamefont {Prokle{\v s}ka}}, \bibinfo {author} {\bibfnamefont {T.}~\bibnamefont {Haidamak}}, \bibinfo {author} {\bibfnamefont {G.}~\bibnamefont {Bastien}}, \bibinfo {author} {\bibfnamefont {V.}~\bibnamefont {Sechovsk{\'y}}}, \bibinfo {author} {\bibfnamefont {G.~G.}\ \bibnamefont {Lonzarich}}, \bibinfo {author} {\bibfnamefont {M.}~\bibnamefont {Vali{\v s}ka}}, \bibinfo {author} {\bibfnamefont {F.~M.}\ \bibnamefont {Grosche}},\ and\ \bibinfo {author} {\bibfnamefont {A.~G.}\ \bibnamefont {Eaton}},\ }\href {https://doi.org/10.1103/PhysRevLett.132.266503} {\bibfield
  {journal} {\bibinfo  {journal} {Physical Review Letters}\ }\textbf {\bibinfo {volume} {132}},\ \bibinfo {pages} {266503} (\bibinfo {year} {2024})}\BibitemShut {NoStop}%
\bibitem [{\citenamefont {Miyake}\ \emph {et~al.}(2019)\citenamefont {Miyake}, \citenamefont {Shimizu}, \citenamefont {Sato}, \citenamefont {Li}, \citenamefont {Nakamura}, \citenamefont {Homma}, \citenamefont {Honda}, \citenamefont {Flouquet}, \citenamefont {Tokunaga},\ and\ \citenamefont {Aoki}}]{Miyake2019a}%
  \BibitemOpen
  \bibfield  {author} {\bibinfo {author} {\bibfnamefont {A.}~\bibnamefont {Miyake}}, \bibinfo {author} {\bibfnamefont {Y.}~\bibnamefont {Shimizu}}, \bibinfo {author} {\bibfnamefont {Y.~J.}\ \bibnamefont {Sato}}, \bibinfo {author} {\bibfnamefont {D.}~\bibnamefont {Li}}, \bibinfo {author} {\bibfnamefont {A.}~\bibnamefont {Nakamura}}, \bibinfo {author} {\bibfnamefont {Y.}~\bibnamefont {Homma}}, \bibinfo {author} {\bibfnamefont {F.}~\bibnamefont {Honda}}, \bibinfo {author} {\bibfnamefont {J.}~\bibnamefont {Flouquet}}, \bibinfo {author} {\bibfnamefont {M.}~\bibnamefont {Tokunaga}},\ and\ \bibinfo {author} {\bibfnamefont {D.}~\bibnamefont {Aoki}},\ }\href {https://doi.org/10.7566/JPSJ.88.063706} {\bibfield  {journal} {\bibinfo  {journal} {Journal of the Physical Society of Japan}\ }\textbf {\bibinfo {volume} {88}},\ \bibinfo {pages} {063706} (\bibinfo {year} {2019})}\BibitemShut {NoStop}%
\bibitem [{\citenamefont {Miyake}\ \emph {et~al.}(2021)\citenamefont {Miyake}, \citenamefont {Shimizu}, \citenamefont {Sato}, \citenamefont {Li}, \citenamefont {Nakamura}, \citenamefont {Homma}, \citenamefont {Honda}, \citenamefont {Flouquet}, \citenamefont {Tokunaga},\ and\ \citenamefont {Aoki}}]{Miyake2021}%
  \BibitemOpen
  \bibfield  {author} {\bibinfo {author} {\bibfnamefont {A.}~\bibnamefont {Miyake}}, \bibinfo {author} {\bibfnamefont {Y.}~\bibnamefont {Shimizu}}, \bibinfo {author} {\bibfnamefont {Y.~J.}\ \bibnamefont {Sato}}, \bibinfo {author} {\bibfnamefont {D.}~\bibnamefont {Li}}, \bibinfo {author} {\bibfnamefont {A.}~\bibnamefont {Nakamura}}, \bibinfo {author} {\bibfnamefont {Y.}~\bibnamefont {Homma}}, \bibinfo {author} {\bibfnamefont {F.}~\bibnamefont {Honda}}, \bibinfo {author} {\bibfnamefont {J.}~\bibnamefont {Flouquet}}, \bibinfo {author} {\bibfnamefont {M.}~\bibnamefont {Tokunaga}},\ and\ \bibinfo {author} {\bibfnamefont {D.}~\bibnamefont {Aoki}},\ }\href {https://doi.org/10.7566/JPSJ.90.103702} {\bibfield  {journal} {\bibinfo  {journal} {Journal of the Physical Society of Japan}\ }\textbf {\bibinfo {volume} {90}},\ \bibinfo {pages} {103702} (\bibinfo {year} {2021})}\BibitemShut {NoStop}%
\bibitem [{\citenamefont {Niu}\ \emph {et~al.}(2020)\citenamefont {Niu}, \citenamefont {Knebel}, \citenamefont {Braithwaite}, \citenamefont {Aoki}, \citenamefont {Lapertot}, \citenamefont {Vali{\v s}ka}, \citenamefont {Seyfarth}, \citenamefont {Knafo}, \citenamefont {Helm}, \citenamefont {Brison}, \citenamefont {Flouquet},\ and\ \citenamefont {Pourret}}]{Niu2020}%
  \BibitemOpen
  \bibfield  {author} {\bibinfo {author} {\bibfnamefont {Q.}~\bibnamefont {Niu}}, \bibinfo {author} {\bibfnamefont {G.}~\bibnamefont {Knebel}}, \bibinfo {author} {\bibfnamefont {D.}~\bibnamefont {Braithwaite}}, \bibinfo {author} {\bibfnamefont {D.}~\bibnamefont {Aoki}}, \bibinfo {author} {\bibfnamefont {G.}~\bibnamefont {Lapertot}}, \bibinfo {author} {\bibfnamefont {M.}~\bibnamefont {Vali{\v s}ka}}, \bibinfo {author} {\bibfnamefont {G.}~\bibnamefont {Seyfarth}}, \bibinfo {author} {\bibfnamefont {W.}~\bibnamefont {Knafo}}, \bibinfo {author} {\bibfnamefont {T.}~\bibnamefont {Helm}}, \bibinfo {author} {\bibfnamefont {J.-P.}\ \bibnamefont {Brison}}, \bibinfo {author} {\bibfnamefont {J.}~\bibnamefont {Flouquet}},\ and\ \bibinfo {author} {\bibfnamefont {A.}~\bibnamefont {Pourret}},\ }\href {https://doi.org/10.1103/PhysRevResearch.2.033179} {\bibfield  {journal} {\bibinfo  {journal} {Physical Review Research}\ }\textbf {\bibinfo {volume} {2}},\ \bibinfo {pages} {033179} (\bibinfo {year} {2020})}\BibitemShut
  {NoStop}%
\bibitem [{\citenamefont {Knebel}\ \emph {et~al.}(2024)\citenamefont {Knebel}, \citenamefont {Pourret}, \citenamefont {Rousseau}, \citenamefont {Marquardt}, \citenamefont {Braithwaite}, \citenamefont {Honda}, \citenamefont {Aoki}, \citenamefont {Lapertot}, \citenamefont {Knafo}, \citenamefont {Seyfarth}, \citenamefont {Brison},\ and\ \citenamefont {Flouquet}}]{Knebel2024}%
  \BibitemOpen
  \bibfield  {author} {\bibinfo {author} {\bibfnamefont {G.}~\bibnamefont {Knebel}}, \bibinfo {author} {\bibfnamefont {A.}~\bibnamefont {Pourret}}, \bibinfo {author} {\bibfnamefont {S.}~\bibnamefont {Rousseau}}, \bibinfo {author} {\bibfnamefont {N.}~\bibnamefont {Marquardt}}, \bibinfo {author} {\bibfnamefont {D.}~\bibnamefont {Braithwaite}}, \bibinfo {author} {\bibfnamefont {F.}~\bibnamefont {Honda}}, \bibinfo {author} {\bibfnamefont {D.}~\bibnamefont {Aoki}}, \bibinfo {author} {\bibfnamefont {G.}~\bibnamefont {Lapertot}}, \bibinfo {author} {\bibfnamefont {W.}~\bibnamefont {Knafo}}, \bibinfo {author} {\bibfnamefont {G.}~\bibnamefont {Seyfarth}}, \bibinfo {author} {\bibfnamefont {J.-P.}\ \bibnamefont {Brison}},\ and\ \bibinfo {author} {\bibfnamefont {J.}~\bibnamefont {Flouquet}},\ }\href {https://doi.org/10.1103/PhysRevB.109.155103} {\bibfield  {journal} {\bibinfo  {journal} {Physical Review B}\ }\textbf {\bibinfo {volume} {109}},\ \bibinfo {pages} {155103} (\bibinfo {year} {2024})}\BibitemShut {NoStop}%
\bibitem [{\citenamefont {Salamone}\ \emph {et~al.}(2023)\citenamefont {Salamone}, \citenamefont {Hugdal}, \citenamefont {Jacobsen},\ and\ \citenamefont {Amundsen}}]{Salamone2023}%
  \BibitemOpen
  \bibfield  {author} {\bibinfo {author} {\bibfnamefont {T.}~\bibnamefont {Salamone}}, \bibinfo {author} {\bibfnamefont {H.~G.}\ \bibnamefont {Hugdal}}, \bibinfo {author} {\bibfnamefont {S.~H.}\ \bibnamefont {Jacobsen}},\ and\ \bibinfo {author} {\bibfnamefont {M.}~\bibnamefont {Amundsen}},\ }\href {https://doi.org/10.1103/PhysRevB.107.174516} {\bibfield  {journal} {\bibinfo  {journal} {Physical Review B}\ }\textbf {\bibinfo {volume} {107}},\ \bibinfo {pages} {174516} (\bibinfo {year} {2023})}\BibitemShut {NoStop}%
\bibitem [{\citenamefont {Han}\ and\ \citenamefont {Kivelson}(2022)}]{Han2022}%
  \BibitemOpen
  \bibfield  {author} {\bibinfo {author} {\bibfnamefont {Z.}~\bibnamefont {Han}}\ and\ \bibinfo {author} {\bibfnamefont {S.~A.}\ \bibnamefont {Kivelson}},\ }\href@noop {} {\bibfield  {journal} {\bibinfo  {journal} {Physical Review B}\ }\textbf {\bibinfo {volume} {105}},\ \bibinfo {pages} {L100509} (\bibinfo {year} {2022})}\BibitemShut {NoStop}%
\bibitem [{\citenamefont {Chakraborty}\ and\ \citenamefont {{Black-Schaffer}}(2024)}]{Chakraborty2024}%
  \BibitemOpen
  \bibfield  {author} {\bibinfo {author} {\bibfnamefont {D.}~\bibnamefont {Chakraborty}}\ and\ \bibinfo {author} {\bibfnamefont {A.~M.}\ \bibnamefont {{Black-Schaffer}}},\ }\href {https://doi.org/10.1103/PhysRevB.110.L060508} {\bibfield  {journal} {\bibinfo  {journal} {Physical Review B}\ }\textbf {\bibinfo {volume} {110}},\ \bibinfo {pages} {L060508} (\bibinfo {year} {2024})}\BibitemShut {NoStop}%
\bibitem [{\citenamefont {Clepkens}\ and\ \citenamefont {Kee}(2024)}]{Clepkens2024}%
  \BibitemOpen
  \bibfield  {author} {\bibinfo {author} {\bibfnamefont {J.}~\bibnamefont {Clepkens}}\ and\ \bibinfo {author} {\bibfnamefont {H.-Y.}\ \bibnamefont {Kee}},\ }\href {https://doi.org/10.1103/PhysRevB.109.214512} {\bibfield  {journal} {\bibinfo  {journal} {Physical Review B}\ }\textbf {\bibinfo {volume} {109}},\ \bibinfo {pages} {214512} (\bibinfo {year} {2024})}\BibitemShut {NoStop}%
\bibitem [{\citenamefont {Stoner}(1997)}]{Stoner1997}%
  \BibitemOpen
  \bibfield  {author} {\bibinfo {author} {\bibfnamefont {E.~C.}\ \bibnamefont {Stoner}},\ }\href {https://doi.org/10.1098/rspa.1938.0066} {\bibfield  {journal} {\bibinfo  {journal} {Proceedings of the Royal Society of London. Series A. Mathematical and Physical Sciences}\ }\textbf {\bibinfo {volume} {165}},\ \bibinfo {pages} {372} (\bibinfo {year} {1997})}\BibitemShut {NoStop}%
\bibitem [{\citenamefont {Bardeen}\ \emph {et~al.}(1957)\citenamefont {Bardeen}, \citenamefont {Cooper},\ and\ \citenamefont {Schrieffer}}]{Bardeen1957}%
  \BibitemOpen
  \bibfield  {author} {\bibinfo {author} {\bibfnamefont {J.}~\bibnamefont {Bardeen}}, \bibinfo {author} {\bibfnamefont {L.~N.}\ \bibnamefont {Cooper}},\ and\ \bibinfo {author} {\bibfnamefont {J.~R.}\ \bibnamefont {Schrieffer}},\ }\href {https://doi.org/10.1103/PhysRev.108.1175} {\bibfield  {journal} {\bibinfo  {journal} {Physical Review}\ }\textbf {\bibinfo {volume} {108}},\ \bibinfo {pages} {1175} (\bibinfo {year} {1957})}\BibitemShut {NoStop}%
\bibitem [{\citenamefont {Knafo}\ \emph {et~al.}(2021{\natexlab{b}})\citenamefont {Knafo}, \citenamefont {Knebel}, \citenamefont {Steffens}, \citenamefont {Kaneko}, \citenamefont {Rosuel}, \citenamefont {Brison}, \citenamefont {Flouquet}, \citenamefont {Aoki}, \citenamefont {Lapertot},\ and\ \citenamefont {Raymond}}]{Knafo2021}%
  \BibitemOpen
  \bibfield  {author} {\bibinfo {author} {\bibfnamefont {W.}~\bibnamefont {Knafo}}, \bibinfo {author} {\bibfnamefont {G.}~\bibnamefont {Knebel}}, \bibinfo {author} {\bibfnamefont {P.}~\bibnamefont {Steffens}}, \bibinfo {author} {\bibfnamefont {K.}~\bibnamefont {Kaneko}}, \bibinfo {author} {\bibfnamefont {A.}~\bibnamefont {Rosuel}}, \bibinfo {author} {\bibfnamefont {J.-P.}\ \bibnamefont {Brison}}, \bibinfo {author} {\bibfnamefont {J.}~\bibnamefont {Flouquet}}, \bibinfo {author} {\bibfnamefont {D.}~\bibnamefont {Aoki}}, \bibinfo {author} {\bibfnamefont {G.}~\bibnamefont {Lapertot}},\ and\ \bibinfo {author} {\bibfnamefont {S.}~\bibnamefont {Raymond}},\ }\href {https://doi.org/10.1103/PhysRevB.104.L100409} {\bibfield  {journal} {\bibinfo  {journal} {Physical Review B}\ }\textbf {\bibinfo {volume} {104}},\ \bibinfo {pages} {L100409} (\bibinfo {year} {2021}{\natexlab{b}})}\BibitemShut {NoStop}%
\bibitem [{\citenamefont {Butch}\ \emph {et~al.}(2022)\citenamefont {Butch}, \citenamefont {Ran}, \citenamefont {Saha}, \citenamefont {Neves}, \citenamefont {Zic}, \citenamefont {Paglione}, \citenamefont {Gladchenko}, \citenamefont {Ye},\ and\ \citenamefont {{Rodriguez-Rivera}}}]{Butch2022}%
  \BibitemOpen
  \bibfield  {author} {\bibinfo {author} {\bibfnamefont {N.~P.}\ \bibnamefont {Butch}}, \bibinfo {author} {\bibfnamefont {S.}~\bibnamefont {Ran}}, \bibinfo {author} {\bibfnamefont {S.~R.}\ \bibnamefont {Saha}}, \bibinfo {author} {\bibfnamefont {P.~M.}\ \bibnamefont {Neves}}, \bibinfo {author} {\bibfnamefont {M.~P.}\ \bibnamefont {Zic}}, \bibinfo {author} {\bibfnamefont {J.}~\bibnamefont {Paglione}}, \bibinfo {author} {\bibfnamefont {S.}~\bibnamefont {Gladchenko}}, \bibinfo {author} {\bibfnamefont {Q.}~\bibnamefont {Ye}},\ and\ \bibinfo {author} {\bibfnamefont {J.~A.}\ \bibnamefont {{Rodriguez-Rivera}}},\ }\href {https://doi.org/10.1038/s41535-022-00445-7} {\bibfield  {journal} {\bibinfo  {journal} {npj Quantum Materials}\ }\textbf {\bibinfo {volume} {7}},\ \bibinfo {pages} {1} (\bibinfo {year} {2022})}\BibitemShut {NoStop}%
\bibitem [{\citenamefont {Tokunaga}\ \emph {et~al.}(2023)\citenamefont {Tokunaga}, \citenamefont {Sakai}, \citenamefont {Kambe}, \citenamefont {Opletal}, \citenamefont {Tokiwa}, \citenamefont {Haga}, \citenamefont {Kitagawa}, \citenamefont {Ishida}, \citenamefont {Aoki}, \citenamefont {Knebel}, \citenamefont {Lapertot}, \citenamefont {Kr{\"a}mer},\ and\ \citenamefont {Horvati{\'c}}}]{Tokunaga2023}%
  \BibitemOpen
  \bibfield  {author} {\bibinfo {author} {\bibfnamefont {Y.}~\bibnamefont {Tokunaga}}, \bibinfo {author} {\bibfnamefont {H.}~\bibnamefont {Sakai}}, \bibinfo {author} {\bibfnamefont {S.}~\bibnamefont {Kambe}}, \bibinfo {author} {\bibfnamefont {P.}~\bibnamefont {Opletal}}, \bibinfo {author} {\bibfnamefont {Y.}~\bibnamefont {Tokiwa}}, \bibinfo {author} {\bibfnamefont {Y.}~\bibnamefont {Haga}}, \bibinfo {author} {\bibfnamefont {S.}~\bibnamefont {Kitagawa}}, \bibinfo {author} {\bibfnamefont {K.}~\bibnamefont {Ishida}}, \bibinfo {author} {\bibfnamefont {D.}~\bibnamefont {Aoki}}, \bibinfo {author} {\bibfnamefont {G.}~\bibnamefont {Knebel}}, \bibinfo {author} {\bibfnamefont {G.}~\bibnamefont {Lapertot}}, \bibinfo {author} {\bibfnamefont {S.}~\bibnamefont {Kr{\"a}mer}},\ and\ \bibinfo {author} {\bibfnamefont {M.}~\bibnamefont {Horvati{\'c}}},\ }\href {https://doi.org/10.1103/PhysRevLett.131.226503} {\bibfield  {journal} {\bibinfo  {journal} {Physical Review Letters}\ }\textbf {\bibinfo {volume} {131}},\ \bibinfo
  {pages} {226503} (\bibinfo {year} {2023})}\BibitemShut {NoStop}%
\bibitem [{\citenamefont {Wu}\ \emph {et~al.}(2025{\natexlab{b}})\citenamefont {Wu}, \citenamefont {Chen}, \citenamefont {Weinberger}, \citenamefont {Cabala}, \citenamefont {Graf}, \citenamefont {Skourski}, \citenamefont {Xie}, \citenamefont {Ling}, \citenamefont {Zhu}, \citenamefont {Sechovsk{\'y}}, \citenamefont {Vali{\v s}ka}, \citenamefont {Grosche},\ and\ \citenamefont {Eaton}}]{Wu2025a}%
  \BibitemOpen
  \bibfield  {author} {\bibinfo {author} {\bibfnamefont {Z.}~\bibnamefont {Wu}}, \bibinfo {author} {\bibfnamefont {H.}~\bibnamefont {Chen}}, \bibinfo {author} {\bibfnamefont {T.~I.}\ \bibnamefont {Weinberger}}, \bibinfo {author} {\bibfnamefont {A.}~\bibnamefont {Cabala}}, \bibinfo {author} {\bibfnamefont {D.~E.}\ \bibnamefont {Graf}}, \bibinfo {author} {\bibfnamefont {Y.}~\bibnamefont {Skourski}}, \bibinfo {author} {\bibfnamefont {W.}~\bibnamefont {Xie}}, \bibinfo {author} {\bibfnamefont {Y.}~\bibnamefont {Ling}}, \bibinfo {author} {\bibfnamefont {Z.}~\bibnamefont {Zhu}}, \bibinfo {author} {\bibfnamefont {V.}~\bibnamefont {Sechovsk{\'y}}}, \bibinfo {author} {\bibfnamefont {M.}~\bibnamefont {Vali{\v s}ka}}, \bibinfo {author} {\bibfnamefont {F.~M.}\ \bibnamefont {Grosche}},\ and\ \bibinfo {author} {\bibfnamefont {A.~G.}\ \bibnamefont {Eaton}},\ }\href {https://doi.org/10.1073/pnas.2422156122} {\bibfield  {journal} {\bibinfo  {journal} {Proceedings of the National Academy of Sciences}\ }\textbf {\bibinfo
  {volume} {122}},\ \bibinfo {pages} {e2422156122} (\bibinfo {year} {2025}{\natexlab{b}})}\BibitemShut {NoStop}%
\bibitem [{\citenamefont {Ran}\ \emph {et~al.}(2021)\citenamefont {Ran}, \citenamefont {Saha}, \citenamefont {Liu}, \citenamefont {Graf}, \citenamefont {Paglione},\ and\ \citenamefont {Butch}}]{Ran2021}%
  \BibitemOpen
  \bibfield  {author} {\bibinfo {author} {\bibfnamefont {S.}~\bibnamefont {Ran}}, \bibinfo {author} {\bibfnamefont {S.~R.}\ \bibnamefont {Saha}}, \bibinfo {author} {\bibfnamefont {I.-L.}\ \bibnamefont {Liu}}, \bibinfo {author} {\bibfnamefont {D.}~\bibnamefont {Graf}}, \bibinfo {author} {\bibfnamefont {J.}~\bibnamefont {Paglione}},\ and\ \bibinfo {author} {\bibfnamefont {N.~P.}\ \bibnamefont {Butch}},\ }\href {https://doi.org/10.1038/s41535-021-00376-9} {\bibfield  {journal} {\bibinfo  {journal} {npj Quantum Materials}\ }\textbf {\bibinfo {volume} {6}},\ \bibinfo {pages} {1} (\bibinfo {year} {2021})}\BibitemShut {NoStop}%
\bibitem [{\citenamefont {Virtanen}\ \emph {et~al.}(2020)\citenamefont {Virtanen}, \citenamefont {Gommers}, \citenamefont {Oliphant}, \citenamefont {Haberland}, \citenamefont {Reddy}, \citenamefont {Cournapeau}, \citenamefont {Burovski}, \citenamefont {Peterson}, \citenamefont {Weckesser}, \citenamefont {Bright}, \citenamefont {{van der Walt}}, \citenamefont {Brett}, \citenamefont {Wilson}, \citenamefont {Millman}, \citenamefont {Mayorov}, \citenamefont {Nelson}, \citenamefont {Jones}, \citenamefont {Kern}, \citenamefont {Larson}, \citenamefont {Carey}, \citenamefont {Polat}, \citenamefont {Feng}, \citenamefont {Moore}, \citenamefont {{VanderPlas}}, \citenamefont {Laxalde}, \citenamefont {Perktold}, \citenamefont {Cimrman}, \citenamefont {Henriksen}, \citenamefont {Quintero}, \citenamefont {Harris}, \citenamefont {Archibald}, \citenamefont {Ribeiro}, \citenamefont {Pedregosa}, \citenamefont {{van Mulbregt}},\ and\ \citenamefont {{SciPy 1.0 Contributors}}}]{Scipy2020}%
  \BibitemOpen
  \bibfield  {author} {\bibinfo {author} {\bibfnamefont {P.}~\bibnamefont {Virtanen}}, \bibinfo {author} {\bibfnamefont {R.}~\bibnamefont {Gommers}}, \bibinfo {author} {\bibfnamefont {T.~E.}\ \bibnamefont {Oliphant}}, \bibinfo {author} {\bibfnamefont {M.}~\bibnamefont {Haberland}}, \bibinfo {author} {\bibfnamefont {T.}~\bibnamefont {Reddy}}, \bibinfo {author} {\bibfnamefont {D.}~\bibnamefont {Cournapeau}}, \bibinfo {author} {\bibfnamefont {E.}~\bibnamefont {Burovski}}, \bibinfo {author} {\bibfnamefont {P.}~\bibnamefont {Peterson}}, \bibinfo {author} {\bibfnamefont {W.}~\bibnamefont {Weckesser}}, \bibinfo {author} {\bibfnamefont {J.}~\bibnamefont {Bright}}, \bibinfo {author} {\bibfnamefont {S.~J.}\ \bibnamefont {{van der Walt}}}, \bibinfo {author} {\bibfnamefont {M.}~\bibnamefont {Brett}}, \bibinfo {author} {\bibfnamefont {J.}~\bibnamefont {Wilson}}, \bibinfo {author} {\bibfnamefont {K.~J.}\ \bibnamefont {Millman}}, \bibinfo {author} {\bibfnamefont {N.}~\bibnamefont {Mayorov}}, \bibinfo {author} {\bibfnamefont
  {A.~R.~J.}\ \bibnamefont {Nelson}}, \bibinfo {author} {\bibfnamefont {E.}~\bibnamefont {Jones}}, \bibinfo {author} {\bibfnamefont {R.}~\bibnamefont {Kern}}, \bibinfo {author} {\bibfnamefont {E.}~\bibnamefont {Larson}}, \bibinfo {author} {\bibfnamefont {C.~J.}\ \bibnamefont {Carey}}, \bibinfo {author} {\bibfnamefont {{\.I}.}~\bibnamefont {Polat}}, \bibinfo {author} {\bibfnamefont {Y.}~\bibnamefont {Feng}}, \bibinfo {author} {\bibfnamefont {E.~W.}\ \bibnamefont {Moore}}, \bibinfo {author} {\bibfnamefont {J.}~\bibnamefont {{VanderPlas}}}, \bibinfo {author} {\bibfnamefont {D.}~\bibnamefont {Laxalde}}, \bibinfo {author} {\bibfnamefont {J.}~\bibnamefont {Perktold}}, \bibinfo {author} {\bibfnamefont {R.}~\bibnamefont {Cimrman}}, \bibinfo {author} {\bibfnamefont {I.}~\bibnamefont {Henriksen}}, \bibinfo {author} {\bibfnamefont {E.~A.}\ \bibnamefont {Quintero}}, \bibinfo {author} {\bibfnamefont {C.~R.}\ \bibnamefont {Harris}}, \bibinfo {author} {\bibfnamefont {A.~M.}\ \bibnamefont {Archibald}}, \bibinfo {author}
  {\bibfnamefont {A.~H.}\ \bibnamefont {Ribeiro}}, \bibinfo {author} {\bibfnamefont {F.}~\bibnamefont {Pedregosa}}, \bibinfo {author} {\bibfnamefont {P.}~\bibnamefont {{van Mulbregt}}},\ and\ \bibinfo {author} {\bibnamefont {{SciPy 1.0 Contributors}}},\ }\href {https://doi.org/10.1038/s41592-019-0686-2} {\bibfield  {journal} {\bibinfo  {journal} {Nature Methods}\ }\textbf {\bibinfo {volume} {17}},\ \bibinfo {pages} {261} (\bibinfo {year} {2020})}\BibitemShut {NoStop}%
\end{thebibliography}%

\onecolumngrid
\appendix
\setcounter{secnumdepth}{1} 
\renewcommand{\thesection}{\Alph{section}}
\renewcommand{\thefigure}{A\arabic{figure}}
\setcounter{section}{0}
\setcounter{figure}{0}

\section*{APPENDIX} 

\section{Numerical parameters}
\label{app:num-param}
For all of the numerical results presented in the main text, we use the following parameters (unless otherwise stated): $m_{cx} = 1$, $m_{cy} = 1.5$, $m_{cz} = 100$, $m_{fx} = 20$, $m_{fy} = 30$, $m_{fz} = 2000$, $V_0 = 0.7$, $\mu_c = 2$, $\mu_f = 1$. All values are presented in units where $2m_{cx} a^2 = 1$ ($a$ is a microscopic length scale). Numerical results labeled ``no SOC" have $\gamma_n = 0$, and ``with SOC" have $\gamma_x = \gamma_y = 0.02$, $\gamma_z = 0.01$. We work at zero-temperature ($T=0$).

The mean-field results are obtained with $U_f = 1$ and $\tilde{g} = g/4 = 1$. We include a small Hubbard repulsion $U_c = 0.001$ on the $c$ fermions to stabilize numerical minimization.

\section{Hamiltonian and gap structures in matrix form}
\label{app:ham-matrix}
The Hamiltonian for the normal state can also be compactly written in matrix form over space of spin and species indices. The matrices $\tau_i$ and $\sigma_i$ are Pauli matrices over species ($c$ or $f$) and spin ($\uparrow$ and $\downarrow$) respectively.  
\begin{equation}
H_{0} = \sum_k \vec{\psi}_k^\dagger \left( H_\epsilon + H_Z + H_{hyb} +H_{SOC}\right)\vec{\psi}_k
\end{equation}
where $\vec{\psi}_k = \begin{pmatrix}
\psi_{kc\uparrow} &
\psi_{kc\downarrow} &
\psi_{kf\uparrow} &
\psi_{kf\downarrow}
\end{pmatrix}^T$ is the vector of fermionic annihilation operators and \begin{align}
H_\epsilon &= \text{diag}\left( \epsilon_{c}(\vec{k}),    \epsilon_{c}(\vec{k}) , \epsilon_{f}(\vec{k}) ,  \epsilon_{f}(\vec{k}) \right) \\
H_Z &= -\mathbf{h} \cdot (\tau_0\otimes \pmb{\sigma}) \\ \nonumber
&= -h \big(\sin\theta\cos\phi\cdot  \tau_0\otimes \sigma_x +  \cos\theta \cdot \tau_0\otimes\sigma_y  + \sin\theta\sin\phi \cdot \tau_0\otimes \sigma_z \big) \\ 
H_{hyb} &= -\delta \cdot \tau_x\otimes\sigma_0 \\
H_{SOC} &= \sum_{n} \gamma_{n} w_n(k_x,k_y, k_z) \tau_y\otimes \sigma_n. 
\end{align}

Using this notation, we define the gap matrices $D_i$ (spin quantized along $z$ axis):
\begin{align}
D_1^{\tau\tau',\sigma\sigma'} &= \frac{i}{2} \tau_x^{\tau\tau'}\otimes \sigma_y^{\sigma\sigma'}  \\
D_2^{\tau\tau',\sigma\sigma'}  &= \frac{i}{2}  \tau_y^{\tau\tau'}\otimes \sigma_z^{\sigma\sigma'}  \\
D_3^{\tau\tau',\sigma\sigma'}  &=  \frac{i}{2} \tau_y^{\tau\tau'}\otimes \mathbf{1}^{\sigma\sigma'} \\
D_4^{\tau\tau',\sigma\sigma'}  &=  \frac{i}{2} \tau_y^{\tau\tau'}\otimes\sigma_x^{\sigma\sigma'} 
\label{eq:D-def}
\end{align}

The gap magnitude corresponding to each gap structure is:
\begin{align}
\Delta_1 &= g\sum_k \expval{f_{-k\uparrow}c_{k\downarrow} -  f_{-k\downarrow}c_{k\uparrow} }\\
\Delta_2 &=  g\sum_k \expval{f_{-k\uparrow}c_{k\uparrow} -  f_{-k\downarrow}c_{k\downarrow}} \\
\Delta_3  &=  g \sum_k\expval{ f_{-k\uparrow}c_{k\uparrow} + f_{-k\downarrow}c_{k\downarrow}  }\\
\Delta_4 &=  g\sum_k \expval{ f_{-k\uparrow}c_{k\downarrow} +  f_{-k\downarrow}c_{k\uparrow}}
\label{eq:gap-mag}
\end{align}

\section{Finite-$q$ susceptibility}
\label{app:finite-q}
We discuss the possibility of finite-momentum pairing in  our model. Given that the $q=0$ pairing in our model is suppressed by a sufficiently large field and that it is not a weak-coupling instability, it is natural to ask whether a different type of superconducting order with $q>0$ might be favorable.

\begin{figure}[h]
\begin{centering}
\includegraphics{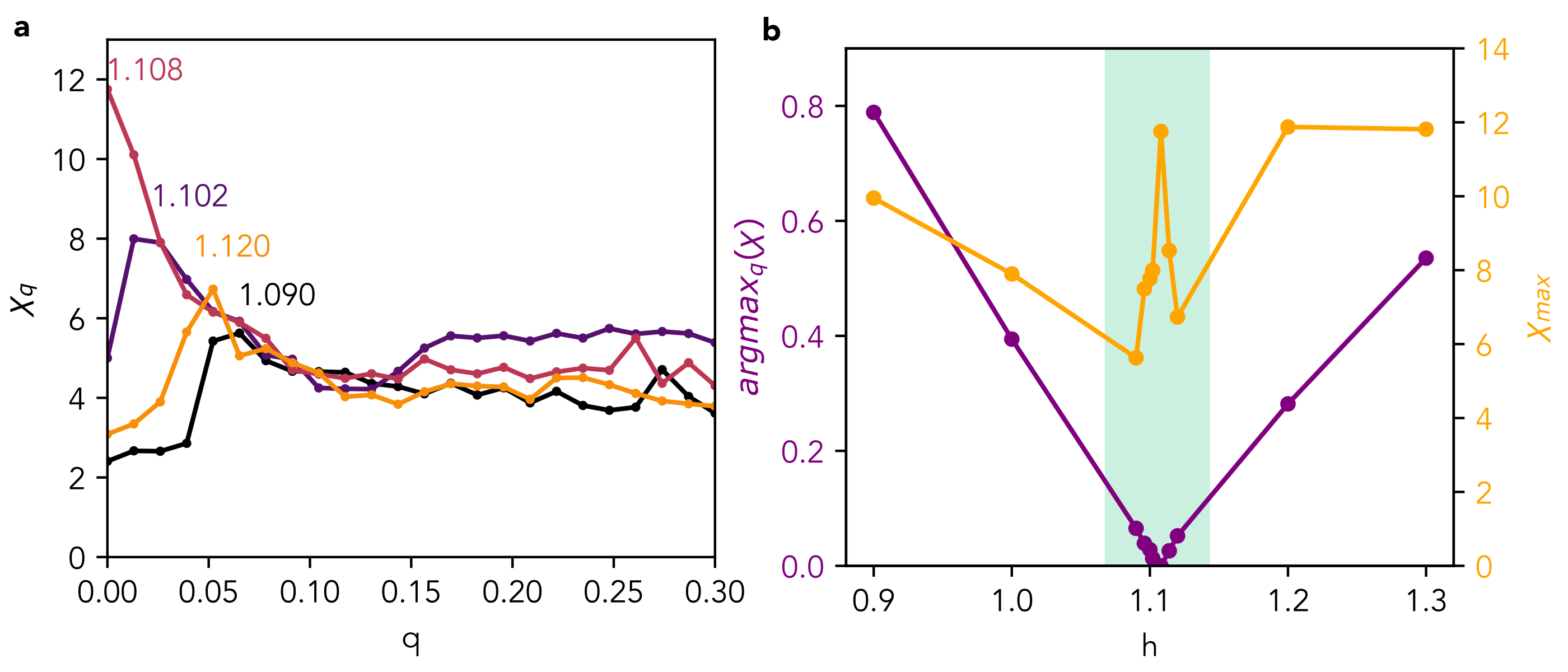}
\end{centering}
\caption{Finite-momentum pair susceptibility $\chi(q)$ when SOC is absent as  (a)  a function $q$ at $\phi=\pi/6, \theta=7\pi/2$ for a range of field strengths. The susceptibility peaks strongly at $q=0$ at a field strength of $h=1.108$. Panel(b) shows $\text{argmax}[\chi(q)]$ (left axis) and $\max \chi(q)$ as a function of the field strength $h$ in a range around $h=1.108$. The shaded region in (b) indicates the range of field strengths displayed in (a). }
\label{suppfig:fin-q}
\end{figure}

We investigate this via the pair susceptibility as a function of the center-of-mass momentum $q$, with the direction fixed to be along $\hat{k_x}$ (Fig. \ref{suppfig:fin-q}). The results for $\mathbf{q}\parallel \hat{k_y}$ are qualitatively similar. 

\begin{figure}[h]
\begin{centering}
\includegraphics{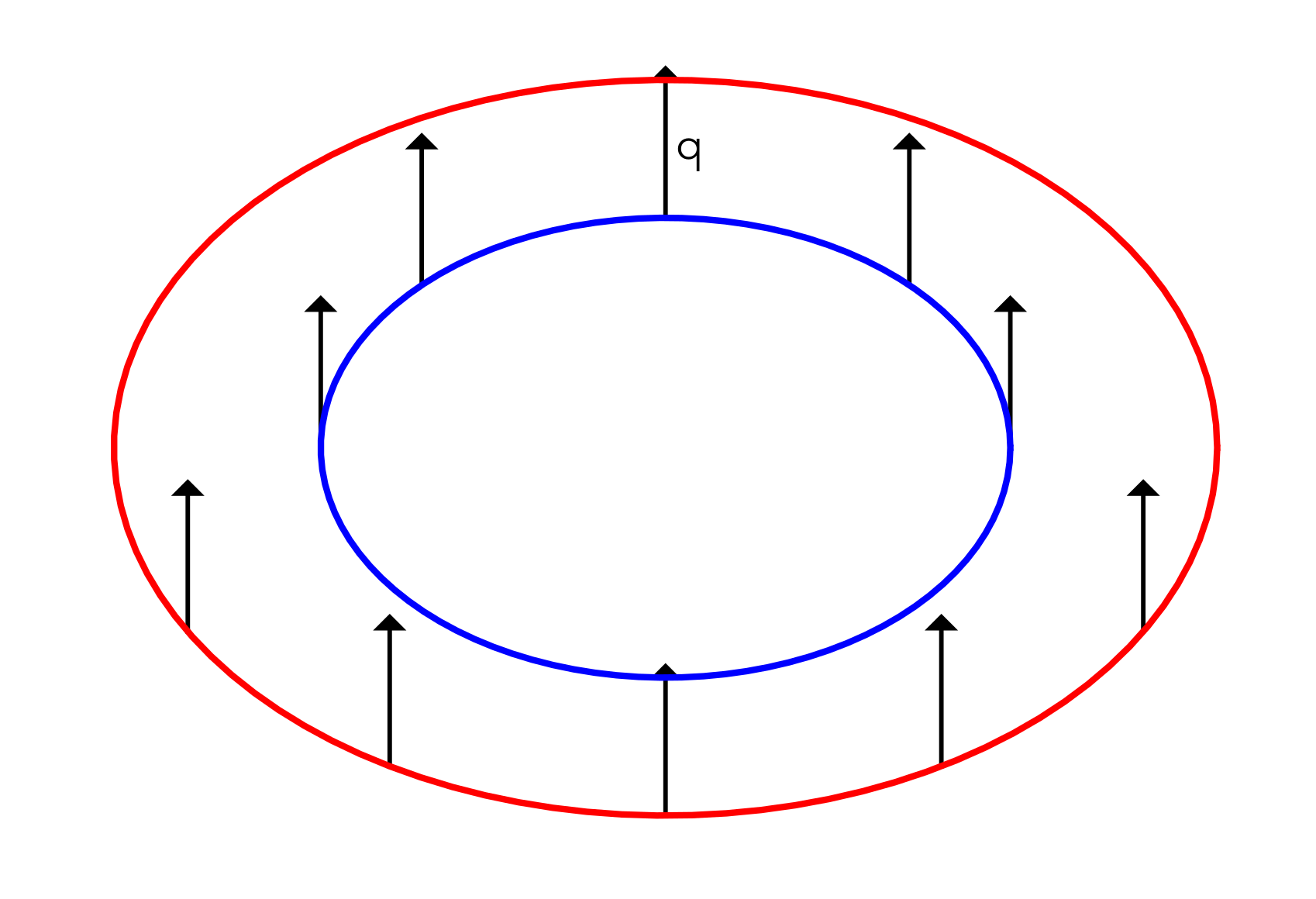}
\end{centering}
\caption{Schematic of how finite-$q$ pairing occurs in a 2D slice of the Fermi surfaces for $\mathbf{q}\parallel \hat{k_y}$. A fixed $q$ connects only certain points on the Fermi surfaces. This imperfect nesting makes the finite-$q$ pairing more fragile than the $q=0$ pairing.}
\label{suppfig:q-nest}
\end{figure}

The fact that the susceptibility peaks at $q=0$ in a range of field strengths can be understood as a consequence of the imperfect nesting at finite-$q$. This is illustrated in Fig.\ref{suppfig:q-nest}

 We highlight that the $q=0$ state, which is the focus of our main text, appears more robust in a certain range of field strengths than finite-$q$ order only for small spin-orbit coupling. However, since superconductivity in our model is not a weak-coupling instability (except in certain fine-tuned cases), this discussion of the pair susceptibility does not constitute a comprehensive study of possible pair-density waves (PDW) in our model. Generically, the stability of finite-$q$ order, especially in comparison to $q=0$ order, will depend on the specific interactions present. 

\section{Orientation-dependence of susceptibility }
\label{app:chi-orientation}
\begin{figure}[h]
\begin{centering}
\includegraphics{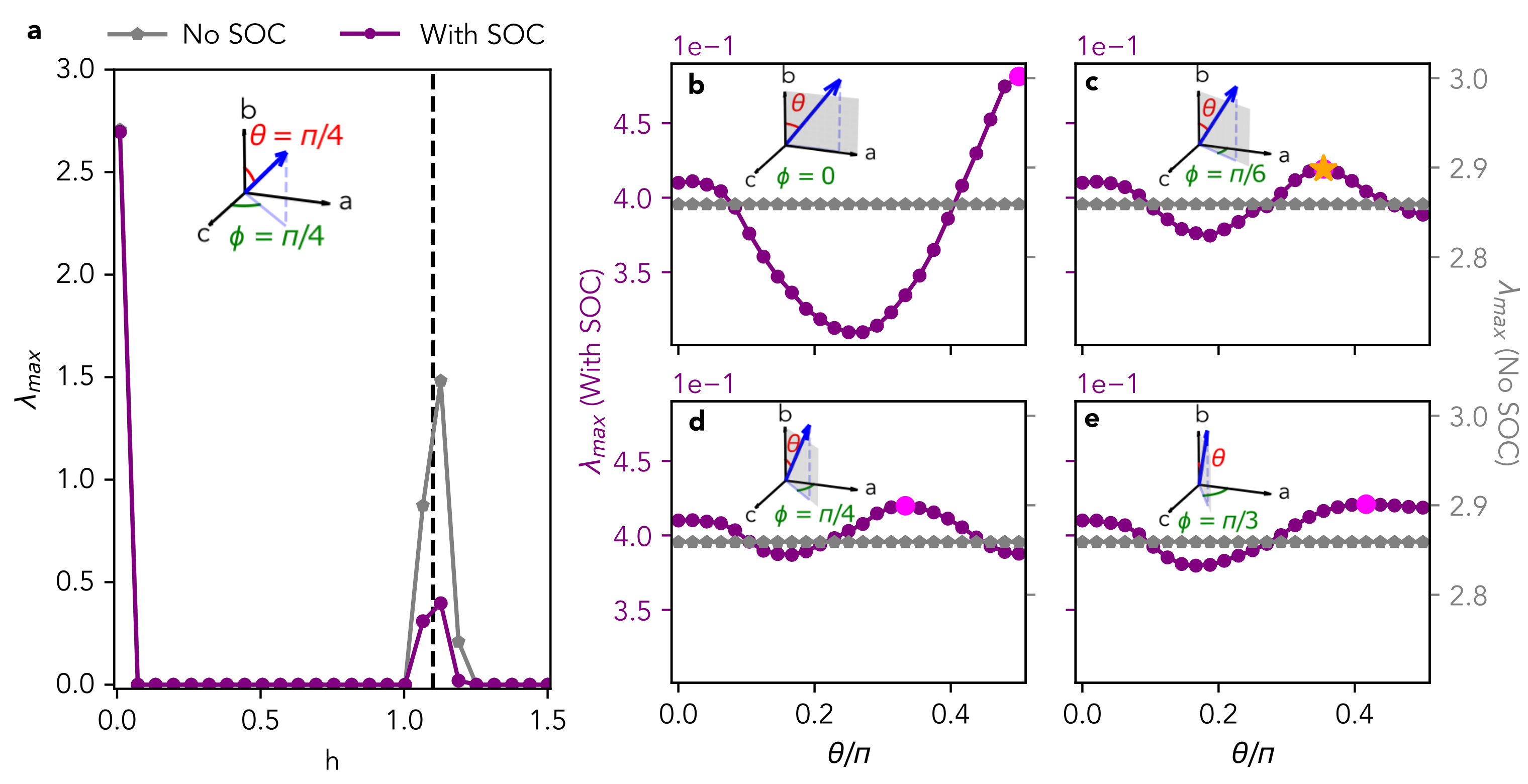}
\end{centering}
\caption{Extended version of Fig. \ref{fig:susc} in the main text. Largest eigenvalue $\lambda_{max}$ of the pair susceptibility matrix $\chi$ as a function of: (a) field magnitude $h$ for the field at a fixed orientation of $\phi=\theta=\pi/4$ and (b-e) angle $\theta$ for fixed magnitude ($h=1.1$) and $\phi= 0, \pi/6, \pi/4, \pi/3$. Purple circles indicate data with SOC, grey pentagons indicate data without SOC. The orange star indicates the largest value of $\lambda_{max}$ in (b); larger magneta circles indicate the largest value of $\lambda_{max}$ in (c), (d), and (e).}
\label{suppfig:chi-orientation-full}
\end{figure}

Here, we include extra data demonstrating the orientation-dependence of the susceptibility in Fig.\ref{suppfig:chi-orientation-full}. 

\section{Pairing mechanisms and gap structures}
\label{app:gap-struct}
While we present numerical results for the case of phonon-mediated attraction in the main text, magnetic interactions could just as well mediate attraction. Here, we tabulate a few possible forms of the interaction and the gaps that they favor. In particular, we focus on possible attraction from: ferromagnetic interactions $H_{FM}$, antiferromagnetic interactions $H_{AFM}$, density-density interactions $H_{d}$ (a generalization of the interaction considered in the main text), and transverse spin interactions $H_{T}$. For simplicity, we will quantize spin along the direction of the field (call this $z$):

\begin{align}
    H_{FM} &= -\sum_{\alpha\beta}\sum_{\mathbf{r}_0, \mathbf{R}} J_{FM}^{\alpha\beta}(\mathbf{R}) S^z_{\mathbf{r}_0,\alpha} S^z_{\mathbf{r}_0+\mathbf{R},\beta}   \\
    H_{AFM} &=  \sum_{\alpha\beta} \sum_{\mathbf{r}_0, \mathbf{R}} J_{AFM}^{\alpha\beta}(\mathbf{R}) S^z_{\mathbf{r}_0,\alpha} S^z_{\mathbf{r}_0+\mathbf{R},\beta}   \\
    H_{d} &=  -\sum_{\alpha\beta}\sum_{\mathbf{r}_0, \mathbf{R}} g^{\alpha\beta}(\mathbf{R}) n_{\mathbf{r}_0,\alpha} n_{\mathbf{r}_0+\mathbf{R}, \beta} \\
    H_T&= -\sum_{\alpha\beta}\sum_{\mathbf{r}_0, \mathbf{R}} J_T^{\alpha\beta}(\mathbf{R}) \left( S^x_{\mathbf{r}_0,,\alpha} S^x_{\mathbf{r}_0+\mathbf{R},\beta}+  S^y_{\mathbf{r}_0,,\alpha} S^y_{\mathbf{r}_0+\mathbf{R},\beta} \right)
\end{align}
where $\alpha$ and $\beta$ are species indices (take values $c$ or $f$), and $J_{FM}, J_{AFM}, g$, and $J_T$ are all positive. Note that we are considering (anti)ferromagnetic interactions along the direction of the field $z$ and transverse spin interactions perpendicular to the field. We assume translational invariance, so the interaction strengths are functions of the relative coordinate $\mathbf{R}$. 

To decouple into different superconducting channels, we Fourier transform and focus only on the zero center-of-mass momentum ($q=0$) terms. The result will look generically like
\begin{equation}
    H_{int} = -\sum_{\alpha\beta\gamma\delta}\sum_k \psi^\dagger_{k\alpha} \psi^\dagger_{-k\beta} \sum_{k'} V^{\alpha\beta\gamma\delta}(\mathbf{k} - \mathbf{k}') \psi_{-\mathbf{k}' \gamma} \psi_{\mathbf{k}' \delta},
\end{equation}
and the gap matrix will be 
\begin{align}
    \Delta^{\alpha\beta}(k) &= \sum_{\gamma\delta}\sum_{k'} V^{\alpha\beta\gamma\delta}(\mathbf{k} - \mathbf{k}') \expval{\psi_{-\mathbf{k}' \gamma} \psi_{\mathbf{k}' \delta}}.
\end{align}
The interaction $V^{\alpha\beta\gamma\delta}(\mathbf{k} - \mathbf{k}')$ can be decomposed into separable pieces which are even or odd in $\mathbf{k}'$ and $\mathbf{k}$. Note that in the main text, we consider only the local ($\mathbf{R} = 0$) part of $H_d$; this corresponds to a momentum-independent interaction. There are now four general cases to consider: the interaction can have components even or odd in $\mathbf{k}'$, and the interaction can be between same or opposite species. 

First, we consider a momentum-even interaction between opposite species ($\alpha\neq \beta$). The gap structures favored by the different interactions and their total associated interaction energies are shown in Table. \ref{tab:mtm-even-opp}. 
\begin{table}[h!]
    \centering
    \begin{tabular}{c||c|c|c|c||c}
       Gap Structure & FM & AFM & Density & Transverse & Total Interaction Energy\\
      \hline
        $D_1 \sim \tau_x\otimes i\sigma_y$ &  + & - & - & + & $ J_{FM}- J_{AFM}-g +J_T $ \\ 
       $D_2\sim i\tau_y\otimes \sigma_x(i\sigma_y)$ &  - & + & - & 0 & $ - J_{FM}+J_{AFM}-g$ \\ 
       $D_3\sim i\tau_y\otimes \sigma_y(i\sigma_y)$ &  - & + & - & 0 &$- J_{FM}+J_{AFM}-g$ \\ 
       $D_4\sim i\tau_y\otimes \sigma_z(i\sigma_y)$ & + & - & - & - & $ J_{FM} - J_{AFM} -g-J_T$ 
    \end{tabular}
    \caption{Momentum-even, opposite species. Negative means attractive, positive means repulsive, and 0 means does not couple to that channel.}
    \label{tab:mtm-even-opp}
\end{table}

From this table, it is clear that the interaction used in the main text is attractive in all channels $D_i$; up to factors of $1/2$, the gap structures above are the same as the $D_i$ previously defined. Consider now the two-band model described in the main text, with a field along the $z$ direction. While all of the gap structures are supported by a local, opposite-species density-density interaction, the accidental crossings occur only for opposite-spin bands, meaning that $D_2$ and $D_3$ do not appear because they are disfavored by the kinetics (rather than by interactions). Because of this, density-density interactions effectively induce the same gaps ($D_1$ and $D_4$) as antiferromagnetic interactions (along the direction of the applied field) in our model. We would also obtain qualitatively similar results from the transverse spin interactions (perpendicular to the applied field), since this is attractive in $D_4$. The gap structure $D_1$ and $D_4$ will generically be supported as inter-band gaps, regardless of the presence of SOC.

For any same-species interactions, the results will depend on the relationship between the $c$ and $f$ interactions (ie., whether $J_{FM}^{cc} = J_{FM}^{ff}$). We assume for simplicity that these are the same. The gap structures for momentum-even, same-species are shown in Table. \ref{tab:mtm-even-same}. 
\begin{table}[h!]
    \centering
    \begin{tabular}{c||c|c|c|c||c}
       Gap Structure & FM & AFM & Density & Transverse & Total Interaction Energy\\
      \hline
       $\tau_0\otimes i\sigma_y$& + & - &- & +&  $J_{FM} - J_{AFM} -g+ J_T$
    \end{tabular}
    \caption{Momentum-even, same species.}
    \label{tab:mtm-even-same}
\end{table}

The only allowed gap structure is the ``conventional" spin-singlet structure. In our two-band model, this is Pauli-limited as usual and would not be field-induced.

For the momentum-odd, opposite-species case, the gap structures and interaction energies are  shown in Table. \ref{tab:mtm-odd-opp}.
\begin{table}[h!]
    \centering
    \begin{tabular}{c||c|c|c|c||c}
       Gap Structure & FM & AFM & Density & Transverse & Total Interaction Energy\\
      \hline
       $i\tau_y\otimes i\sigma_y$ & - & + & + & - &  $-J_{FM}+J_{AFM}+g -J_T $\\
        $\tau_x\otimes \sigma_x(i\sigma_y)$ & - & + & -& 0  & $ -J_{FM}+ J_{AFM}-g $ \\ 
       $\tau_x\otimes \sigma_y(i\sigma_y)$ & - & + & - & 0  & $ -J_{FM}+ J_{AFM}-g $ \\ 
       $\tau_x\otimes \sigma_z(i\sigma_y)$ & + & -& - & - &  $ J_{FM}- J_{AFM}-g  -J_T $
    \end{tabular}
    \caption{Momentum-odd, opposite species. }
    \label{tab:mtm-odd-opp}
\end{table}
Qualitatively, the pairing which is favored by both the interactions and kinetics (ie., which bands are close to the Fermi level) is similar to the momentum-even, opposite-species case: opposite spin and opposite species. Note that this again corresponds to inter-band pairing.

The momentum-odd, same-species case is shown in Table. \ref{tab:mtm-odd-same}.
\begin{table}[h!]
    \centering
    \begin{tabular}{c||c|c|c|c||c}
       Gap Structure & FM & AFM & Density & Transverse & Total Interaction Energy\\
      \hline
       $\tau_0\otimes \sigma_x(i\sigma_y)$& - & + & - & 0 & $-J_{FM}+J_{AFM} - g $  \\
      $\tau_0\otimes \sigma_y(i\sigma_y)$& - & + & - & 0 &  $-J_{FM}+J_{AFM} - g $\\ 
      $\tau_0\otimes \sigma_z(i\sigma_y)$& + & - & - & - & $J_{FM}-J_{AFM} - g - J_T $ 
    \end{tabular}
    \caption{Momentum-odd, same species. }
    \label{tab:mtm-odd-same}
\end{table}
This final set of possibilities includes intra-band, odd-parity, spin-triplet pairing ($\tau_0\otimes \sigma_x(i\sigma_y)$ and $\tau_0\otimes \sigma_y(i\sigma_y)$). 

\section{Mean-field decomposition and free energy}
\label{app:MFT}
We define the mean fields as: 
\begin{align}
    \mathbf{M}_\alpha &= \expval{\mathbf{S}}_\alpha\\
    \Delta_i &= \tilde{g}\sum_{\mathbf{k}} D_i^{\tau\tau',\sigma\sigma'}\expval{\psi_{\mathbf{k}\tau\sigma}\psi_{-\mathbf{k}\tau'\sigma'}}
\end{align}
where we introduce $\tilde{g} = g/4$

 Decoupling the interactions in the mean fields, we find:
\begin{align}
H_{U,MF} &= -\sum_\alpha \frac{U_\alpha}{2}\mathbf{M}_\alpha  \cdot \mathbf{S}_\alpha  + \sum_\alpha \frac{U_\alpha}{4} (M_\alpha^2 + \bar{n}_\alpha^2) \\
H_{g,MF} &= -\frac{1}{2}\sum_{i,k,a,b}\left(c^\dagger_{ka} (D_i^\dagger)^{ab}  c^\dagger_{-kb} \Delta_i + \Delta_i^* c_{-kb} D_i^{ba}c_{ka} \right) + \sum_i\frac{ \Delta_i^2}{2\tilde{g}} .
\end{align}
where $\bar{n}_\alpha$ is the average particle density of species $\alpha$ (ie., density at zero-field).  
We define
\begin{align}
    H_{pair} &=-\frac{1}{2}\sum_{i,k,a,b}\left(c^\dagger_{ka} (D_i^\dagger)^{ab}  c^\dagger_{-kb} \Delta_i + \Delta_i^* c_{-kb} D_i^{ba}c_{ka} \right)  \\
    H_{Z,\text{eff}} &= -\sum_\alpha\left(\mathbf{h} + \frac{U_\alpha}{2} \mathbf{M}_\alpha\right)\cdot (P_\alpha\otimes \pmb{\sigma})
\end{align}
where $P_\alpha$ projects to the species $\alpha$:
\begin{align}
P_c &= \begin{pmatrix}
1 & 0 \\
0 & 0 
\end{pmatrix} \quad \quad P_f = \begin{pmatrix}
0 & 0 \\
0 & 1 
\end{pmatrix}. 
\end{align}

With these mean-field decompositions, the full mean-field Hamiltonian is then
\begin{align}
H_{MF} &= \sum_k \vec{\psi}_k^\dagger ( H_\epsilon + H_{SOC} + H_{hyb} + H_{Z,\text{eff}}  ) \psi_k + H_{pair} + E_0 \\ 
E_0 &=  \sum_\alpha \frac{U_\alpha}{4} (\bar{n}_\alpha^2 + M_\alpha^2)   +\frac{|\Delta|^2}{2\tilde{g}}
\end{align}

To compute the free energy, we expand our basis to both hole and particle space, such that 
\begin{align}
H_{MF} =  \sum_k \Psi^\dagger_k H_{BdG}(k) \Psi_k + E_0 + \sum_{k\alpha\sigma} \tfrac{1}{2} (\epsilon_{\alpha} (\mathbf{k}) -h_{\alpha,\text{eff}}^z)
\end{align}

We diagonalize the Bogoliubov-de Gennes Hamiltonian $H_{BdG}$:
\begin{align}
H_{BdG}(\mathbf{k}) &= \frac{1}{2} \begin{pmatrix}
H_\epsilon(\mathbf{k})+H_{Z,\text{eff}}(\mathbf{k})+H_{hyb}(\mathbf{k})+H_{SOC}(\mathbf{k}) & \hat{\Delta}(\mathbf{k})\\ 
\hat{\Delta}^\dagger(\mathbf{k}) & -(H_\epsilon(-\mathbf{k})+H_{Z,\text{eff}}(-\mathbf{k})+H_{hyb}(-\mathbf{k})+H_{SOC}(-\mathbf{k})) 
\end{pmatrix}
\end{align}
From this, 
\begin{align}
H_{MF} &=  2\sum_k E_{kn} a^\dagger_{kn} a_{kn} + \sum_{k\in k_F, \alpha\sigma} \tfrac{1}{2} (\epsilon_{k\alpha}-h_{\alpha,\text{eff}}^z) - \sum_{k} E_{kn}
\end{align}
where $n$ labels the bands (1, 2, 3, 4). 

The BdG quasiparticles are all noninteracting, so we can construct the free energy using the fermionic many-body partition function for independent particles.  From $Z_{MF} = \exp(-\beta H_{MF})$ and $F_{MF} = -k_B T \ln Z_{MF}$, we have
\begin{align}
F_{MF}= E_0 - k_B T \sum_{kn} \ln \left( 1+ e^{-2\beta E_{kn}}\right) + \left(  \sum_{k\in k_F,\alpha}  \tfrac{1}{2} (\epsilon_{k\alpha}-h_{\alpha,\text{eff}}^z) - \sum_{k\in k_F,n} E_{kn}\right)
\end{align}
For numerical efficiency, it is advantageous to write the middle term as an entropy contribution (which vanishes when $T=0$) and a temperature-independent internal energy contribution: 
\begin{align}
F_{MF}&= k_B T\sum_{kn} \left( f(2E_{kn})  \ln(f(2E_{kn})) + (1-f) \ln(1-f)  \right) \\ \nonumber
&\quad + \sum_{kn \alpha} f(2E_{kn}) 2 E_{kn} + E_0 + \left(  \sum_{k\in k_F,n} \tfrac{1}{2} (\epsilon_{k\alpha}-h_{\alpha,\text{eff}}^z) - \sum_{k\in k_F,n} E_{kn}\right)
\end{align}
where $f(E)$ is the Fermi-Dirac distribution (here, $\mu=0$ for the BdG quasiparticles).

\section{Numerical minimization of free energy}
\label{app:minimization}
We use SciPy\cite{Scipy2020} to minimize the free energy as a function of the gap magnitudes $\Delta_i$ and magnetizations $\mathbf{M}_\alpha$. When constructing the free energy function, we enforce that the gap magnitude in every channel is zero outside of a cutoff energy.  The parameter-space is 14-dimensional: 6 magnetic order parameters ($\mathbf{M}_c$ and $\mathbf{M}_f$, each with three spatial components) and  4  (coefficients of) superconducting order parameters ($\Delta_i$,  with $i=1,2,3,4$, each with a real and imaginary part).  We can reduce this to 13 parameters by choosing the global phase of the superconductivity such that a particular channel always has a purely real gap coefficient.

Note that the gap is restricted to be a linear combination of the forms $D_i$ to have correct antisymmetric properties under fermion exchange. To plot the gap magnitude (as in Fig.\ref{fig:mf}), we rotate to the band basis and take the maximum gap along the Fermi surface. For the numerical parameters, we have chosen, the gap dependence on $\vec{k}$ is relatively weak; this is because SOC is small, and hybridization is $k$-independent, so the transformation to the band basis imparts very little $k$-dependence to the gap.

\section{Mean-field solutions for the gaps}
\label{app:gap-nonunitary}
In the main text, we present only the gap magnitude as a function of field strength. Here, we discuss the form of the gap solution; we find that the gap is non-unitary at finite field. To illustrate this, we describe some representative results for $\tilde{g} = g/4 = 1$ and $U_f = 1$ and a field pointed in the direction $\phi=\pi/3$, $\theta = 0.35\pi$ (the same parameters used to generate Fig. \ref{fig:mf}). 

A unitary gap function obeys $\Delta^\dagger \Delta \propto \mathbf{1}$ (where $\mathbf{1}$ is the identity matrix). Note that this is a basis-independent definition. At zero-field, the solution to the free energy minimization is: $(\Delta_1, \Delta_2, \Delta_3, \Delta_4) = (3.8\times 10^{-3}, -4.5\times 10^{-7}+9\times 10^{-8}i ,4.1\times 10^{-7}-7.2\times 10^{-7}i  ,4.5\times 10^{-7}+7\times 10^{-8}i )$. Using the gap structures $D_i$ previously defined, one can confirm that this is a unitary state:
\begin{align}
     \Delta^\dagger \Delta = \begin{pmatrix}
        1.45\times 10^{-5} &  0 & 0 & 0 \\ 
        0 & 1.45\times 10^{-5}& 0 &0 \\
         0 & 0 &1.45\times 10^{-5} & 0 \\ 
         0& 0& 0 &1.45\times 10^{-5}
    \end{pmatrix}.
\end{align}

In contrast, the solution to the free energy minimization at field $h=0.82$ is: $(\Delta_1, \Delta_2, \Delta_3, \Delta_4) = (-2.62\times10^{-3},  2.28\times10^{-3}+1.64\times10^{-4}i,
  9.7\times 10^{-7}-2.61\times10^{-5}i, -1.56\times10^{-3}+2.37\times 10^{-4}i)$. In this case, $\Delta^\dagger \Delta$ is not proportional to the identity and instead takes the form: 
\begin{align}
    \Delta^\dagger \Delta = \begin{pmatrix}
        2.28\times 10^{-5} &  1.19\times 10^{-5} +1.73\times 10^{-6}i & 0 & 0 \\ 
         1.19\times 10^{-5} -1.73\times 10^{-6}i &  6.42\times 10^{-6} & 0 &0 \\
         0 & 0 &6.41\times 10^{-6} & -1.2 \times 10^{-5} +1.46\times 10^{-5}i \\ 
         0& 0& -1.2 \times 10^{-5} -1.46\times 10^{-5}i & 2.28 \times 10^{-5}
    \end{pmatrix}.
\end{align}
It is unsurprising (and in fact expected) that, in the presence of a large field, the resulting gap matrix should be non-unitary. This is because a unitary gap preserves time-reversal symmetry (TRS). Since the magnetic field breaks TRS, a large field could induce a gap which also breaks TRS. 

Note that, when minimizing the free energy, we simultaneously obtain solutions for the gap coefficients and magnetizations. In the above, our focus is the non-unitary gap structure, so we omit the solutions for magnetization for clarity. 

\section{Mean-field solutions: other orientations}
\label{app:MFT-orientation}
In the main text, we present mean-field solutions for magnetization and the gap magnitude for a field in the direction $\phi=\pi/6$, $\theta=7\pi/20$. Here, we include some extra data for the field pointing in another direction to demonstrate the SOC-induced anisotropy. 
\begin{figure}
\begin{centering}
\includegraphics{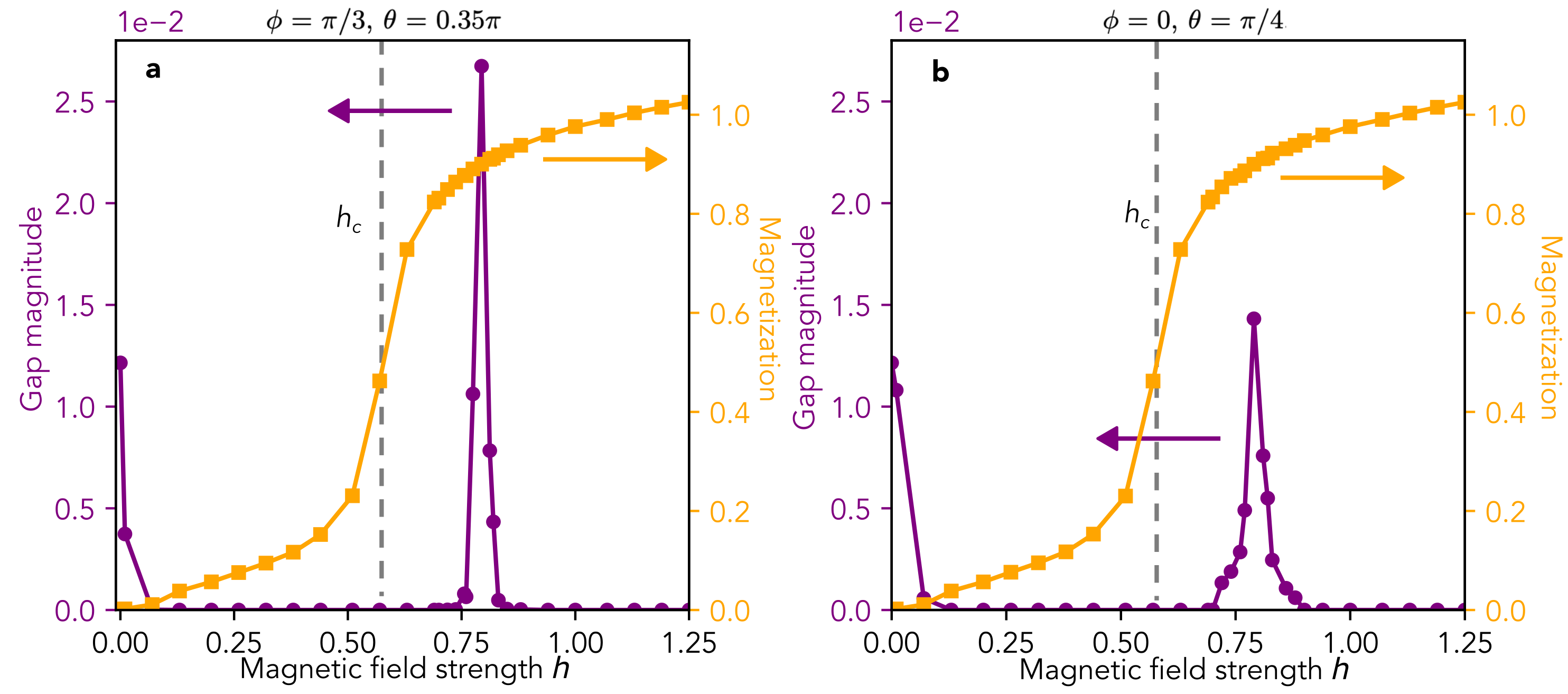}
\end{centering}
\caption{ Results of minimizing the free energy of the mean-field Hamiltonianwith $g=4$, $U_f=1$ for a magnetic field $h$ in the orientation, shown for (a) $\phi=\pi/3$, $\theta = 0.35\pi$ (same as Fig. \ref{fig:mf} in the main text, star in Fig. \ref{fig:susc}c) and (b)  $\phi=0$, $\theta = \pi/4$. Each plot shows both the maximum gap magnitude (left axis) total magnetization (right axis) at each magnetic field strength. Critical field for the metamagnetic transition $h_c$ is marked by a dashed line.}
\label{suppfig:mf-tilt}
\end{figure}
As shown in Fig.\ref{suppfig:mf-tilt}, the gap magnitude induced by the magnetic field depends on the orientation of the field. This is consistent from the field-dependence suggested by our analysis of the pair susceptibility. 

\section{Mean-field solutions: critical fields for metamagnetism and superconductivity}
\label{app:MFT-relative-field}
\begin{figure}[h]
\begin{centering}
\includegraphics{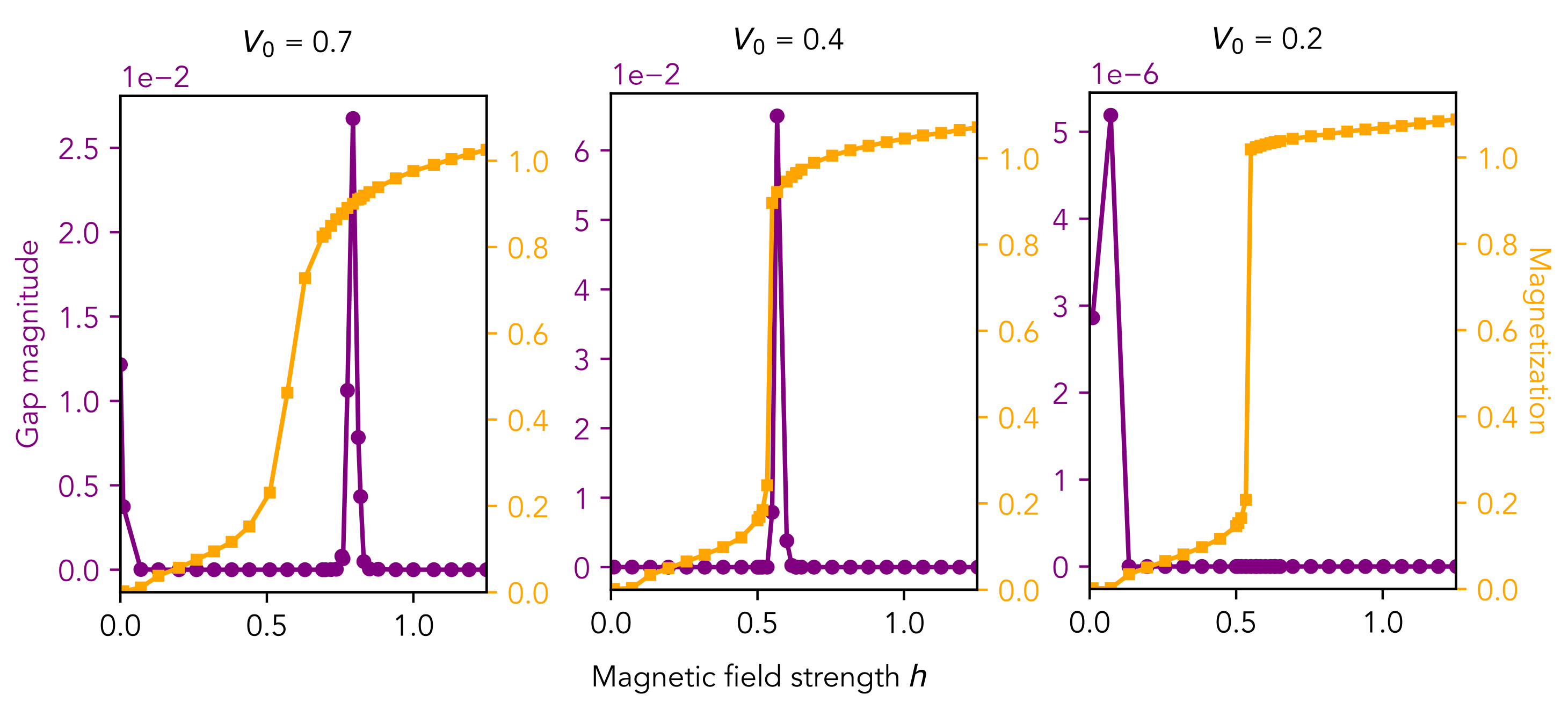}
\end{centering}
\caption{Results of minimizing free energy of the mean-field Hamiltonian with $g=4$, $U_f=1$ as a function of magnetic field strength $h$ in the direction $\phi=\pi/6$, $\theta=7\pi/20$ for different hybridizations: (a) $V_0= 0.7$ (as in main text Fig. \ref{fig:mf}), (b)$V_0= 0.4$, and (c) $V_0= 0.2$. In (b), the region of superconductivity coincides with the metamagnetic transition, in contrast to (a), where the metamagnetic transition occurs at a lower field strength than the superconductivity. In (c), superconductivity is suppressed for this choice of parameters, as the jump in magnetization at the metamagnetic transition causes the system to avoid band crossings at $E_F$. }
\label{suppfig:mft-hyb}
\end{figure}

We also include extra data for different hybridization strengths to demonstrate how the field strength at which superconductivity occurs is not fixed with relation to the critical field strength of the metamagnetic transition (Fig. \ref{suppfig:mft-hyb}). In this case, we illustrate this by tuning the hybridization, but note that changes in the masses, interactions, and relative chemical potentials will also influence the quantitative results. 

A smaller hybridization produces in quasiparticle bands that more strongly resemble the original $c$ and $f$ bands; as a result, the $f$-like quasiparticle band with lower Fermi velocity is less dispersive for smaller hyrbridization, causing a sharper metamagnetic transition at the same interaction strength. 

\section{Odd-parity pairing}
\label{app:odd-parity}
We elaborate here on the possibility of odd-parity pairing. If we consider odd-parity pairing on both the itininerant and heavy bands, superconductivity would persist at all field strengths. If instead superconductivity has support only on the heavy band, the state still avoids conventional Pauli limiting, but superconductivity exists only when the heavy band crosses $E_F$; thus, depopulation of the heavy band (rather than de-pairing) leads to an upper critical field. We will focus on the latter case. We imagine, for example, ferromagnetic nearest-neighbor interactions which are strong on the $f$ quasiparticles and weak on the $c$ quasiparticles. These favor same-spin pairing on the heavy band. 

Let us refer to the field strength at which superconductivity is centered as $h_{SC}$ and the field strength around which the metamagnetic transition is centered as $h_{c}$. Then, our claim is that $h_{SC} \approx h_c$ for odd-parity pairing on the heavy band. We highlight that this statement is in contrast to the even-parity case, in which the relationship between $h_{SC}$ and $h_c$ can be tuned by the band parameters (as shown in the previous section). 

\begin{figure}[h]
\begin{centering}
\includegraphics{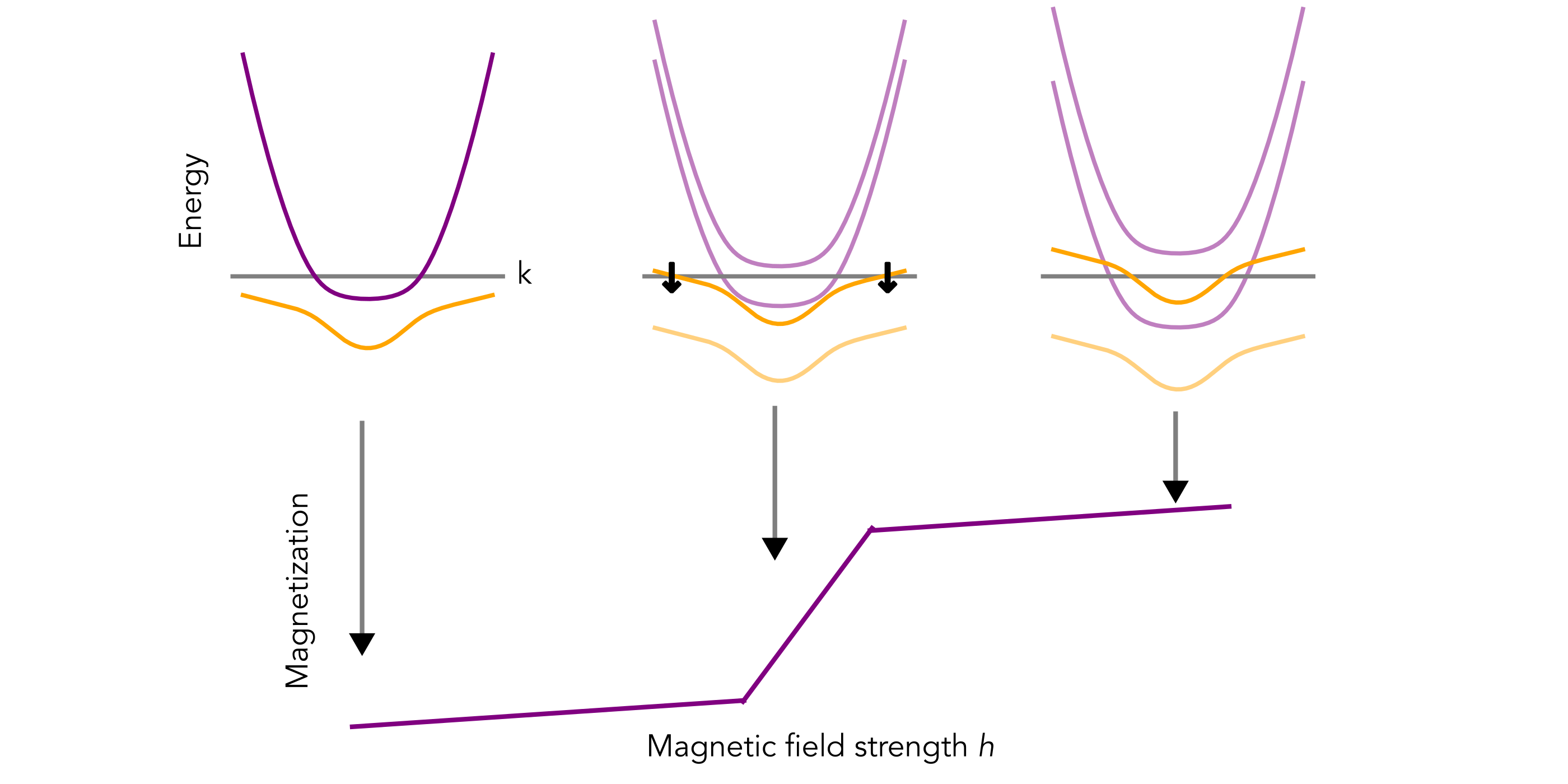}
\end{centering}
\caption{ Schematic of magnetization as a function of magnetic field and accompanying band energies. The middle panel shows how odd-parity superconductivity in the $f$ species must arise at field strengths within coinciding with the metamagnetic transition (the steep increase in magnetization). }
\label{suppfig:odd-parity}
\end{figure}

The metamagnetic transition can be understood as a consequence of field-induced depopulation of the heavy $f$ band; the beginning and end of depopulation correspond to the beginning and end of the metamagnetic transition. This is manifest as a region of abrupt and dramatic increase in the magnetization. As the interaction $U_f$ is increased, the transition sharpens but generically still occurs over a finite range of magnetic field strengths. For superconductivity supported on the heavy band, pairing necessarily occurs only if the $f$ band crosses $E_F$; otherwise, there are no particles to pair. Thus, superconductivity necessarily occurs at field strengths contained within the range over which the metamagnetic transition occurs. Fig. \ref{suppfig:odd-parity} shows an exaggerated schematic for this argument. 

In summary, odd-parity superconductivity on the $f$ band is still field-induced and occurs in a range of field strengths away from 0. However, within our model, it does not persist for field strengths outside of the metamagnetic transition range.

\end{document}